\documentclass[%
 reprint,
 amsmath,amssymb,
]{revtex4-2}

\usepackage{graphicx}
\usepackage{dcolumn}
\usepackage{bm}
\usepackage{enumitem}

\usepackage{physics}
\usepackage{braket}
\usepackage{hyperref}
\usepackage{ulem}
\usepackage{float}
\usepackage{here}
\usepackage{lineno}
\usepackage{cancel}

\usepackage{color}

\usepackage{amsmath}
\usepackage{amssymb, amsfonts}
\begin{document}
\twocolumngrid 
\preprint{APS/123-QED}

\title{Zepto to attosecond core-level photoemission time delays in homonuclear diatomic molecules and non-dipole effects in the framework of Multiple Scattering theory}

\author{
Yoshiaki Tamura$^{1*}$,
Kaoru Yamazaki$^{2}$,
Kiyoshi Ueda$^{3,4}$ and
Keisuke Hatada$^{1}$
}
\email{Authors to whom any corresponding should be addressed\\E-mail: hatada@sci.u-toyama.ac.jp,\\yoshiakitamura.susa@gmail.com}
\affiliation{$^1$ Department of Physics, University of Toyama, 3190 Gofuku, 930-8555 Toyama, Japan}
\affiliation{$^2$ Attosecond Science Research Team, Extreme Photonics Research Group, RIKEN Center for Advanced Photonics, RIKEN, 2-1 Hirosawa, Wako, 351-0198 Saitama, Japan.}
\affiliation{$^3$ Department of Chemistry, Graduate School of Science, Tohoku University, 6-3 Aramaki Aza-Aoba, Aoba-ku, 980-8578 Sendai, Japan}
\affiliation{$^4$ School Physical Science and Technology, ShanghaiTech University, Shanghai 201210, P. R. China}






\date{\today}

\begin{abstract}
This study theoretically investigates the angular distribution of core-level photoemission time delay within a molecular frame.
This phenomenon can be measured with the advancement of attosecond pulsed lasers and metrology.
Our focus is on homonuclear diatomic molecules.
The two-center interference patterns observed in the \textit{gerade} and \textit{ungerade} core-level Molecular-Frame Photoelectron Angular Distributions (MFPAD) of homonuclear diatomic molecules demonstrate symmetry breaking with respect to the direction of light propagation, attributed to the non-dipole (multipole) effect.
This non-dipole effect reflects the zeptosecond time discrepancy in the initiation of photoelectron waves at the two absorbing atoms.
Our study delves into the photoemission time delay resulting from the non-dipole effect through the introduction of a theoretical model.
We reveal that when considering the contributions from the \textit{gerade} and \textit{ungerade} delocalized states in incoherent sums, the two-center interference terms cancel each other in both the MFPADs and photoemission time delays.
However, a residual term persists showcasing the non-dipole effect in the photoemission time delays.
Furthermore, by expanding the scattering state of photoelectrons using the Multiple Scattering theory, we demonstrate the significant role played by the scattering of photoelectrons at the molecular potential in describing the photoemission time delays of homonuclear diatomic molecules.
Additionally, we establish that the direct wave contribution does not exhibit angular dependence in the photoemission time delays.
Next, we apply our theoretical model to a nitrogen molecule, demonstrating the energy- and angular-dependent characteristics of the MFPADs and photoemission time delays through both analytical and numerical approaches.
The incoherent sums of the MFPADs in both forward and backward directions exhibit equal intensity, whereas the incoherent sums of the photoemission time delays show a slight variation of a few hundred zeptoseconds compared with numerical calculations using a multiple scattering code.
Our findings demonstrate that the photoemission time delay serves as a valuable tool for investigating attosecond and zeptosecond photoionization dynamics, providing insights that may be overlooked in the MFPAD analysis.
This study provides a basis for theoretical investigations into photoionization phenomena in nano systems containing numerous identical atoms, such as large polyatomic molecules and condensed matter, enabling the examination of photoemission time delays and non-dipole effects.
\end{abstract}

\maketitle


\section {Introduction}

In this decade, attosecond pulsed lasers have revolutionized the experimental study of electron dynamics on the attosecond scale~\cite{Krausz2009, Nisoli2017, BorregoVarillas2022}.
A key inquiry addressed by cutting-edge attosecond metrology is the precise measurement of photoionization timescales, a topic that garnered attention in the 2023 Nobel Prize in Physics~\cite {Nobel}.
The photoemission time delay reflects atto to zeptosecond ($10^{-18}$–$10^{-21}$ s.) electron dynamics, including its phase, \textit{e.g.}, zeptosecond photon-electron interaction~\cite{Amusia2020,Liang2024}, electron excitation~\cite{Schultze2010, Vos2018, Gong2022}, attosecond scattering, and trapping with surrounding potentials during electron propagation in matter~\cite{Kotur2016, Gruson2016, Kaldun2016, Huppert2016, Cirelli2018, Loriot2020, Nandi2020, Heck2021, Gong2022, Ahmadi2022}, as well as subsequent escape and hole localization~\cite{Gong2022_2}.
These dynamics are largely independent of the slower femtosecond structural dynamics.
Photoemission time delay measurements offer valuable real-time insights into previously unobservable ultrafast phenomena and provide attosecond probes into various chemical reaction dynamics.

As proof-of-principle experiments, photoemission time delay has been successfully measured in the extreme ultraviolet region for atoms~\cite{Schultze2010, Kotur2016, Gruson2016, Kaldun2016, Isinger2017_new, Azoury2018, Cirelli2018, Peschel2022}, molecules~\cite{Huppert2016, Beaulieu2017, Vos2018, Loriot2020, Nandi2020, Kamalov2020, Heck2021, Gong2022, Ahmadi2022, Loriot2024}, clusters~\cite{Gong2022_2}, and condensed phases~\cite{Cavalieri2007_new, Jordan2020_new}.
Theoretical predictions have indicated that the photoemission angle dependence of molecular photoemission time delay can be attributed to the anisotropy of molecular potentials~\cite{Ivanov2012, Serov2013, Chacon2014, Hockett2016, baykusheva2017, Serov2017, biswas2020}.
Recent advancements in technology, such as the combination of Cold Target Recoil Ion Momentum Spectroscopy with attosecond pulse trains recently realized the angle-resolved measurement of photoemission time delay in the molecular frame~\cite{Vos2018, Heck2021, Gong2022, Ahmadi2022}.
Furthermore, the use of soft X-ray attosecond pulses has allowed for the measurement of photoemission time delay of 1$s$ core-level electrons~\cite{Driver_Nature,Ji_arXiv}, with potential applications in solid and liquid phases.

Numerous studies have explored the angular dependence of photoelectron emission from core orbitals of diatomic molecules, providing a solid basis for their extension to more complex molecular systems.
In molecules with localized core orbitals, such as heteronuclear diatomic molecules, the core-level Molecular Frame Photoelectron Angular Distribution (MFPAD) shows interference patterns due to multiple scattering in the molecular potential~\cite{Ota2021}.
In contrast, in homonuclear diatomic molecules with non-localized core orbitals, the core-level MFPAD exhibits characteristic oscillation patterns~\cite{Nagy2004, Liu2006, Akoury2007, Hara2022} related to Cohen-Fano two-center interference~\cite {Cohen1966}, in addition to multiple scattering.
The photoemission time delay in such two-center systems has been both theoretically~\cite{Ivanov2012, Serov2013} and experimentally~\cite{Heck2022} reported within the Electric Dipole approximation.

Recently, a remarkable symmetry breaking with respect to the direction of light propagation on the oscillation pattern owing to two-center interference in MFPAD of hydrogen molecules was experimentally observed with high-precision measurements at high photon energy
~\cite{Grundmann2020, Rezvan2024}.
This symmetry breaking originates from a non-dipole (multipole) effect ignored in the Electric Dipole model and is observed in the photon energy range above $\sim 500$ eV~\cite{Hemmers2001, Langhoff2001}.
Such non-dipole effect can be interpreted as a zeptosecond time difference in the birth of photoemission waves at the two light-absorbing atoms owing to light propagation between them~\cite{Ivanov2021}.

The aforementioned non-dipole effects, originating from equivalent atoms, may significantly influence photoelectron angular distributions in various systems.
However, these effects have not been thoroughly examined.
Furthermore, the study of photoemission time delay may offer valuable insights that cannot be obtained through angle-resolved photoelectron spectroscopy alone.
This includes the phase information of the non-dipole transition amplitudes, as demonstrated in previous research on atoms~\cite{Liang2024,Amusia2020}.
Homonuclear diatomic molecules, comprising two identical atoms, represent the smallest system for study.
Understanding the non-dipole effect on core-level MFPAD and molecular-frame photoemission time delay angler distribution in homonuclear diatomic molecules is crucial for expanding our knowledge of these nano systems, as highlighted in research on valence photoemission time delay in polyaromatic hydrocarbons~\cite{Loriot2024}.

This study presents a theoretical model for core-level photoemission time delay in homonuclear diatomic molecules.
The model incorporates single-channel, one-photon, and one-electron ionization from core levels, as well as an approximation beyond the Electric Dipole approximation.
Moreover, we analytically determined the essential contributions of photoelectron scattering by molecular potentials to the two-center interference and non-dipole effect in the photoemission time delay and derive analytical expressions.
Our comparison of calculations based on these analytical formulas with numerical calculations for the nitrogen molecule provides new insights into the non-dipole effects of core-level photoemission from two equivalent atoms.

In summary, our model accurately reproduces the backward-forward tilt of the oscillation pattern of two-center interference observed in MFPAD for both \textit{gerade} and \textit{ungerade} states, as previously reported~\cite{Grundmann2020, Ivanov2021}.
The corresponding photoemission time delay reached an extreme value at emission angles where the MFPAD approached zero, exhibiting forward and backward asymmetries.
Notably, when combining the \textit{gerade} and \textit{ungerade} states incoherently, as in most experiments, the forward and backward asymmetries in the MFPAD along the direction of light propagation disappear.
However, the asymmetry in the photoemission time delay owing to non-dipole effects persists.
Our analysis reveals that the non-dipole photoemission time delay contains information regarding a zeptosecond-order time delay associated with electron dynamics immediately following light irradiation, which cannot be deduced from MFPADs.

\section{Theory}

\subsection{Beyond Electric Dipole approximation describing interatomic light propagation in homonuclear diatomic molecules}

Grundmann \textit{et al.} and Ivanov \textit{et al.} reported that the multipole effect related to light propagation between two atoms is crucial in homonuclear diatomic molecules.~\cite{Grundmann2020, Ivanov2021}
This multipole effect is effectively incorporated into the photoionization amplitude as a phase difference in the light propagation between two atoms by utilizing the Electric Dipole ($E1$) approximation solely for transitions within atoms, which is hereafter referred to as the $E1'$ approximation.
This approximation is valid in scenarios where the core orbital's dispersion is smaller than the bond length, such as the nitrogen molecules (refer to Appendix~\ref{derivation_E1'} for details).
\begin{align}
&
    \bra{\,\psi^{-}(\mathbf{k})\,}\,
    e^{\mathrm{i}\boldsymbol{\kappa}\cdot\mathbf{r}}\,\,
    \hat{\boldsymbol{\varepsilon}}
    \cdot 
    \mathbf{r}\,
    \ket{\,\phi^{\,g/u}}
    \nonumber\\
&
    \xrightarrow[\substack{
    \mathrm{E1'\,\,approximation}
    }]{}
    \nonumber\\
&
    \frac{1}{\sqrt{2}}\,\,
    \Bigl(
    e^{-\mathrm{i}\boldsymbol{\kappa}\cdot\frac{\mathbf{R}}{2}}\,\,
    A_{1}
    (E1; \mathbf{k}, \hat{\boldsymbol{\varepsilon}})
    \pm
    e^{\mathrm{i}\boldsymbol{\kappa}\cdot\frac{\mathbf{R}}{2}}\,\,
    A_{2}
    (E1; \mathbf{k}, \hat{\boldsymbol{\varepsilon}})
    \Bigr)
    \label{eq:E1'_approximation}
    \\
&   
    \hspace{4.0 cm}
    \equiv
    A^{g/u}
    (E1'; \mathbf{k}, \hat{\boldsymbol{\varepsilon}}, \boldsymbol{\kappa})\,.
    \label{eq:amp_E1'}
\end{align}
Here, we utilized the expression for the photoionization amplitude based on a one-photon and one-electron framework, as well as a single-channel model within the first-order perturbation theory in length gauge~\cite{bethe2012quantum}.
We utilized atomic units for length and Rydberg units for energies ($2m/\hbar^{2}\rightarrow1$; therefore, $k^{2}=E$) throughout the study.
$\ket{\,\psi^{-}(\mathbf{k})\,}$ denotes a photoelectron continuum state with the incoming boundary condition and the momentum vectors $\mathbf{k}$ and $\ket{\,\phi^{\,g/u}}$ denote the \textit{gerade}/\textit{ungerade} states.
$\hat{\boldsymbol{\varepsilon}}$ and $\boldsymbol{\kappa}$ represent the polarization and momentum vectors of the incoming light, respectively.
$\mathbf{R}$ denotes the vector from sites $1$ to $2$, which defines the molecular axis.
In the aforementioned equations,
\begin{align}
    A_{i}
    (E1; \mathbf{k}, \hat{\boldsymbol{\varepsilon}})
=
    \int_{V_{i}}
    \mathrm{d}\mathbf{r}_{i}\,\,
    \psi^{-\,*}_{i}(\mathbf{k}, \mathbf{r}_{i})\,\,
    \hat{\boldsymbol{\varepsilon}}
    \cdot 
    \mathbf{r}_{i}\,\,
    \phi_{i,L_{c}}(r_{i})
    \label{eq:amp_i}
\end{align}
represents the photoionization amplitude, which describes the $E1$ transition from the core orbital of the atom $i$.
$r_{i}=\left|\mathbf{r}_{i}\right|$ on the other hand, represents a norm of a position vector from the center of atom $i$, $\psi^{-}_{i}$ represents the continuum state wave function defined only in the atomic sphere $i$, including multiple scattering contributions by surrounding atoms, $\phi_{i,L_{c}}(r_{i})$ represents a core wave function of atom $i$ with angular momentum quantum number $L_{c}\equiv(l_{c},m_{c})=(0,0)$ and the volume integrals, $\int_{V_{i}}d\mathbf{r}$, are conducted within the spherical region of site $i$ with radius $b$.

For simplicity, our theoretical model is based on one-electron and single-channel models.
In this model, we assumed that only one molecular orbital in the initial state participates in the photoemission process and the wave function of the photoelectron in the final state can be specified by a unique momentum vector $\mathbf{k}$.
In the framework of the many-body Keldysh Green's function formalism, $\psi^{-}(\mathbf{k},\mathbf{r})$ and $\phi^{\,g/u}(\mathbf{r})$ on the left-hand side of Eq.~(\ref{eq:E1'_approximation}) should be replaced by particle Dyson orbitals
$f_{n}(x)\equiv\bra{0,N}\hat{\psi}(x)\ket{n,N+1}$ and hole Dyson orbitals
$g_{p}(x)\equiv\bra{p,N-1}\hat{\psi}(x)\ket{0,N}$
respectively~\cite{Almbladh_and_Hedin_book,vonNiessen1984,Fujikawa1999,Fujikawa2005,Ortiz2020}.
$\hat{\psi}(x)$ denote the electron annihilation operators with $x=\left(\mathbf{r},\sigma\right)$, $\sigma$ represents the spin coordinate ($\sigma=$ up, down).
$\ket{n,N+1}$ represents $n$th particle state of a $(N+1)$-electron system.
$\ket{p,N-1}$ represents the $p$th hole state with $\left(N-1\right)$-electrons.
Note that $\mathbf{k}$ is implicitly specified by $p$ and $n$ for a given photon energy.
Based on a previous study of Ref. ~\cite{Fujikawa1999}, the Dyson orbitals can be approximated as $f_{n}(x)\sim\psi^{-}(\mathbf{k},\mathbf{r})$ and $g_{p}(x)\sim S^{g/u}_{0}\,\phi^{\,g/u}(\mathbf{r})$, respectively.
Here, $S^{\,g/u}_{0}=\bra{0,N-1}\hat{c}^{\,g/u}\ket{0,N}$ denotes the loss amplitude and
$\hat{c}^{\,g/u}$ denotes the annihilation operator of the molecular orbitals $\phi^{\,g/u}$ from which an electron gets detached by the photoemission process.
We applied a single-channel model of $S^{\,g/u}_{0}$ and approximated it to $S^{\,g/u}_{0}\sim 1$.
Finally, our model focuses on the \textit{gerade}/\textit{ungerade} symmetry of the scattering states of the photoelectron and that the experimental example in the manuscripts deals with the final state symmetry.

By applying the aforementioned model, the MFPAD from the \textit{gerade} and \textit{ungerade} states can be described as follows:
\begin{align}
&
    I^{g/u}
    (
    \mathbf{k}, 
    \hat{\boldsymbol{\varepsilon}}, 
    \boldsymbol{\kappa}
    )
\equiv
    8\pi^{2}c\kappa
    \left|
    \bra{\,\psi^{-}(\mathbf{k})\,}\,
    e^{\mathrm{i}\boldsymbol{\kappa}\cdot\mathbf{r}}\,
    \hat{\boldsymbol{\varepsilon}}
    \cdot 
    \mathbf{r}\,
    \ket{\,\phi^{\,g/u}}
    \right|^{2}
    \nonumber\\
&
\xrightarrow[\substack{
    \mathrm{E1'\,\,approximation}
    }]{}
    8\pi^{2}c\kappa
    \left|
    A^{g/u}
    (E1'; \mathbf{k}, \hat{\boldsymbol{\varepsilon}}, \boldsymbol{\kappa})
    \right|^{2}
    \nonumber\\
&\hspace{4.5 cm}
    \equiv
    I^{g/u}
    (
    E1'; 
    \mathbf{k}, 
    \hat{\boldsymbol{\varepsilon}}, 
    \boldsymbol{\kappa}
    )
    \,,
    \label{eq:MFPAD_after_E1'}
\end{align}
where
\begin{align}
&
    I^{g/u}
    (
    E1'; 
    \mathbf{k}, 
    \hat{\boldsymbol{\varepsilon}}, 
    \boldsymbol{\kappa}
    )
=
    \frac{1}{2}
    I_{1}
    (E1; \mathbf{k}, \hat{\boldsymbol{\varepsilon}})
    +
    \frac{1}{2}
    I_{2}
    (E1; \mathbf{k}, \hat{\boldsymbol{\varepsilon}})
    \nonumber\\
&\pm
    8\pi^{2}c\kappa\,\,
    \mathrm{Re}
    \Bigl\{
    e^{\mathrm{i}\boldsymbol{\kappa}\cdot\mathbf{R}}\,
    A_{1}^{*}
    (E1; \mathbf{k}, \hat{\boldsymbol{\varepsilon}})\,
    A_{2}
    (E1; \mathbf{k}, \hat{\boldsymbol{\varepsilon}})
    \Bigr\}
    \label{eq:MFPAD_E1'}
\end{align}
and
\begin{align}
    I_{i}
    (E1; \mathbf{k}, \hat{\boldsymbol{\varepsilon}})
\equiv
    8\pi^{2}c\kappa
    \left|\,
    A_{i}
    (E1; \mathbf{k}, \hat{\boldsymbol{\varepsilon}})
    \right|^{2}
\end{align}
represents an MFPAD from the core orbital $\phi_{i,L_{c}}$ localized in atom $i$ of homonuclear diatomic molecules.
The third term in Eq.~(\ref{eq:MFPAD_E1'}) describes the two-center interference phenomenon.

The photoemission time delays in the \textit{gerade} and \textit{ungerade} states are expressed as follows:
\begin{align}
&
    t^{g/u}
    (
    \mathbf{k}, 
    \hat{\boldsymbol{\varepsilon}}, 
    \boldsymbol{\kappa}
    )
\equiv
    \frac{\mathrm{d}}{\mathrm{d}E_{\mathbf{k}}}
    \,
    \mathrm{arg}
    \left\{
    \bra{\,\psi^{-}(\mathbf{k})\,}\,
    e^{\mathrm{i}\boldsymbol{\kappa}\cdot\mathbf{r}}\,
    \hat{\boldsymbol{\varepsilon}}
    \cdot 
    \mathbf{r}\,
    \ket{\,\phi^{\,g/u}}
    \right\}
    \nonumber\\
&\xrightarrow[\substack{
    \mathrm{E1'\,\,approximation}
    }]{}
    \frac{\mathrm{d}}{\mathrm{d}E_{\mathbf{k}}}
    \,
    \mathrm{arg}
    \left\{
    A^{g/u}
    (E1'; \mathbf{k}, \hat{\boldsymbol{\varepsilon}}, \boldsymbol{\kappa})
    \right\}
    \nonumber\\
&\hspace{4.5 cm}
    \equiv
    t^{g/u}
    (
    E1'; 
    \mathbf{k}, 
    \hat{\boldsymbol{\varepsilon}}, 
    \boldsymbol{\kappa}
    )
    \,.
    \label{eq:time_delay_after_E1'}
\end{align}
By utilizing the chain rule of differentiation, the energy derivative can be separated into the following two terms:
\begin{align}
    \frac
    {\mathrm{d}}
    {\mathrm{d}E_{\mathbf{k}}}
=
    \frac
    {\mathrm{d}k}
    {\mathrm{d}E_{\mathbf{k}}}
    \frac
    {\partial}
    {\partial k}
+
    \frac
    {\mathrm{d}\kappa}
    {\mathrm{d}E_{\mathbf{k}}}
    \frac
    {\partial}
    {\partial \kappa}
= 
    \frac
    {1}
    {v_{\mathrm{g}}}
    \frac
    {\partial}
    {\partial k}
+
    \frac{1}{c}
    \frac
    {\partial}
    {\partial \kappa}\,,
    \label{eq:chain_rule}
\end{align}
where $E_{\mathbf{k}}=\left|\mathbf{k}\right|^{2}$ represents photoelectron energy and ${v_{\mathrm{g}}}={\mathrm{d}E_{\mathbf{k}}}/{\mathrm{d}k}=2k$ represents the group veracity of a free electron wave packet with momentum $k=\left|\mathbf{k}\right|$.
From the relationship $\kappa=\left|\boldsymbol{\kappa}\right|=\left(E_{\mathbf{k}}+I_{\mathrm{p}}\right)/c$, where $I_{\mathrm{p}}$ represents the ionization energy and $c$ represents velocity of light, we obtain ${\mathrm{d}\kappa}/{\mathrm{d}E_{\mathbf{k}}}=1/c$.

The photoemission time delay can be separated by applying Eq.~(\ref{eq:chain_rule}):
\begin{align}
    t^{g/u}
    (
    E1'; 
    \mathbf{k}, 
    \hat{\boldsymbol{\varepsilon}}, 
    \boldsymbol{\kappa}
    )
&=
    t_{\partial k}^{g/u}
    (
    E1'; 
    \mathbf{k}, 
    \hat{\boldsymbol{\varepsilon}}, 
    \boldsymbol{\kappa}
    )
+
    t_{\partial \kappa}^{g/u}
    (
    E1'; 
    \mathbf{k}, 
    \hat{\boldsymbol{\varepsilon}}, 
    \boldsymbol{\kappa}
    )
    \,,
    \label{eq:TD_ed_nd}
\end{align}
where
\begin{align}
&   t_{\partial k}^{g/u}
    (E1';
    \mathbf{k}, 
    \hat{\boldsymbol{\varepsilon}}, 
    \boldsymbol{\kappa}
    )
\equiv
    \frac{1}{v_{\mathrm{g}}}
    \frac
    {\partial}
    {\partial k}\,
    \mathrm{arg}
    \left\{
    A^{g/u}
    (E1'; \mathbf{k}, \hat{\boldsymbol{\varepsilon}}, \boldsymbol{\kappa})
    \right\}
    \nonumber\\
&=
    \left(I^{g/u}
    (E1';\mathbf{k}, \hat{\boldsymbol{\varepsilon}}, \boldsymbol{\kappa})\right)^{-1}
    \Biggl[
    I_{1}
    (E1;\mathbf{k}, \hat{\boldsymbol{\varepsilon}})\,
    t_{1}
    (E1; \mathbf{k}, \hat{\boldsymbol{\varepsilon}})
    \nonumber\\
&\hspace{3.75 cm}+
    I_{2}
    (E1;\mathbf{k}, \hat{\boldsymbol{\varepsilon}})\,
    t_{2}
    (E1; \mathbf{k}, \hat{\boldsymbol{\varepsilon}})
    \nonumber\\
&\hspace{0.25 cm}\pm
    \mathrm{Im}
    \Biggl\{
    e^{\mathrm{i}\boldsymbol{\kappa}\cdot\mathbf{R}}\,
    A_{1}^{*}
    (E1; \mathbf{k}, \hat{\boldsymbol{\varepsilon}})\,
    \frac{1}{v_{\mathrm{g}}}
    \frac
    {\partial}
    {\partial k}\,
    A_{2}
    (E1; \mathbf{k}, \hat{\boldsymbol{\varepsilon}})
    \nonumber\\
&\hspace{0.25 cm}\hspace{0.5 cm}+
    e^{-\mathrm{i}\boldsymbol{\kappa}\cdot\mathbf{R}}\,
    A_{2}^{*}
    (E1; \mathbf{k}, \hat{\boldsymbol{\varepsilon}})\,
    \frac{1}{v_{\mathrm{g}}}
    \frac
    {\partial}
    {\partial k}\,
    A_{1}
    (E1; \mathbf{k}, \hat{\boldsymbol{\varepsilon}})
    \Biggr\}
    \Biggr]\,,
    \label{eq:ted_gu}
\end{align}
\begin{align}
&   t_{\partial \kappa}^{g/u}
    (E1';
    \mathbf{k}, 
    \hat{\boldsymbol{\varepsilon}}, 
    \boldsymbol{\kappa}
    )
\equiv
    \frac{1}{c}
    \frac
    {\partial}
    {\partial \kappa}\,
    \mathrm{arg}
    \left\{
    A^{g/u}
    (E1'; \mathbf{k}, \hat{\boldsymbol{\varepsilon}}, \boldsymbol{\kappa})
    \right\}
    \nonumber\\
&=
    \frac{1}{c}
    \frac{R}{2}\,
    \hat{\mathbf{\boldsymbol{\kappa}}}
    \cdot
    \hat{\mathbf{R}}
    \,
    \frac
    {
    I_{2}
    (E1;\mathbf{k}, \hat{\boldsymbol{\varepsilon}})
    -
    I_{1}
    (E1;\mathbf{k}, \hat{\boldsymbol{\varepsilon}})
    }
    {
    2\,I^{g/u}
    (E1';\mathbf{k}, \hat{\boldsymbol{\varepsilon}}, \boldsymbol{\kappa})
    }
    \label{eq:tnd_gu}
\end{align}
and
\begin{align}
    t_{i}
    (E1; \mathbf{k}, \hat{\boldsymbol{\varepsilon}})
\equiv
    \frac{\mathrm{d}}{\mathrm{d}E_{\mathbf{k}}}
    \,
    \mathrm{arg}
    \left\{
    A_{i}
    (E1; \mathbf{k}, \hat{\boldsymbol{\varepsilon}})
    \right\}\,.
\end{align}
Here $t_{\partial k}^{g/u}$ represents the photoemission time delay resulting from the $k$-partial derivative in Eq.~(\ref{eq:chain_rule}).
The third term in the square brackets in Eq.~(\ref{eq:ted_gu}) is responsible for two-center interference phenomenon specific to the photoemission time delay.
Conversely, the first and second terms in Eq.~(\ref{eq:ted_gu}) are characterized by $t_{i}$ which represents the photoemission time delay from the core orbital $\phi_{i,L_{c}}$ localized in an atom $i$ of the homonuclear diatomic molecules.
Further details on $t_{i}$ are provided in~\cite{Tamura2022}.
$t_{\partial \kappa}^{g/u}$ denotes the photoemission time delay corresponding to the $\kappa$-partial derivative in Eq.~(\ref{eq:chain_rule}).
The factor $\left(R/2c\right)\hat{\mathbf{\boldsymbol{\kappa}}}\cdot\hat{\mathbf{R}}$ results from $\kappa$-partial derivative of the factor $\exp\left(\pm\mathrm{i}\boldsymbol{\kappa}\cdot\mathbf{R}/2\right)$, where $R/c\sim366.12$ zs represents the traveling time for light propagation between atoms.
Because both $t_{\partial k}^{g/u}$ and $t_{\partial \kappa}^{g/u}$ are divided by $I^{g/u}$, they exhibit the particular behavior of two-center interference phenomenon specific to MFPAD.

\subsection {Incoherent sums of MFPAD and photoemission time delay for \textit{gerade} and \textit{ungerade} states}
In most experiments, distinguishing photoelectrons from \textit{gerade} and \textit{ungerade} states is technically challenging because the energy splitting between these states is of the sub-eV order (\textit{e.g.}, $97$ meV for nitrogen molecules~\cite {Hergenhahn2001}).
The superposition of photoelectron waves from the \textit{gerade} and \textit{ungerade} states should be performed incoherently~\cite{Cherepkov2008,Schoffler2011}:
\begin{align}   
    \braket{I}_{\mathrm{IC}}
&\equiv
    \frac{1}{2}
    \left(
    I^{g}+I^{u}
    \right)\,.
    \label{eq:IS_MFPAD}
\end{align}
In Ref.~\cite{Schoffler2011}, Sch\"{o}ffler \textit{et al.} validated that the incoherent sum results yielded excellent agreement with all available data on the angular distribution of photoelectrons.~\cite{Cherepkov2000,Jahnke2002}
This observation underscores the orthogonality of the \textit{gerade} and \textit{ungerade}.
Based on a study by Smith~\cite{Smith1960}, the photoemission time delay can be expressed as the following incoherent sum:
\begin{align}
    \braket{t}_{\mathrm{IC}}
&\equiv
    \frac
    {I^{g}\,t^{g}+I^{u}\,t^{u}}
    {I^{g}+I^{u}}\,.
    \label{eq:IS_time_delay}
\end{align}
Heck \textit{et al.}~\cite{Heck2022} utilized this average of the photoemission time delays for krypton dimers and obtained excellent agreement between the theoretical results and experimental data.

For the models described in Eqs.~(\ref{eq:MFPAD_after_E1'}) and~(\ref{eq:time_delay_after_E1'}), the corresponding incoherent sums for the photoemission time delays can be expressed as follows:
\begin{align}
    \braket{I
    (
    E1';
    \mathbf{k}, 
    \hat{\boldsymbol{\varepsilon}}
    )
    }_{\mathrm{IC}}
&=
    \frac{1}{2}
    \left(
    I_{1}
    (E1; \mathbf{k}, \hat{\boldsymbol{\varepsilon}})
    +
    I_{2}
    (E1; \mathbf{k}, \hat{\boldsymbol{\varepsilon}})
    \right)
    \,,
    \label{eq:I_sum}
\\
    \braket{t_{\partial k}
    (
    E1';
    \mathbf{k}, 
    \hat{\boldsymbol{\varepsilon}}
    )
    }_{\mathrm{IC}}
&\nonumber\\
&\hspace{-2.5 cm}=
    \frac{
    I_{1}
    (E1; \mathbf{k}, \hat{\boldsymbol{\varepsilon}})\,
    t_{1}
    (E1; \mathbf{k}, \hat{\boldsymbol{\varepsilon}})
    +
    I_{2}
    (E1; \mathbf{k}, \hat{\boldsymbol{\varepsilon}})\,
    t_{2}
    (E1; \mathbf{k}, \hat{\boldsymbol{\varepsilon}})
    }
    {
    2
    \braket{I
    (
    E1';
    \mathbf{k}, 
    \hat{\boldsymbol{\varepsilon}}
    )
    }_{\mathrm{IC}}
    }
    \,,
    \label{eq:teq_sum}
\\
    \braket{t_{\partial\kappa}
    (
    E1';
    \mathbf{k}, 
    \hat{\boldsymbol{\varepsilon}}, 
    \boldsymbol{\kappa}
    )
    }_{\mathrm{IC}}
&=
    \frac{R}{2c}\,\,
    \hat{\mathbf{\boldsymbol{\kappa}}}
    \cdot
    \hat{\mathbf{R}}\,\,
    \frac
    {
    I_{2}
    (E1; \mathbf{k}, \hat{\boldsymbol{\varepsilon}})
    -
    I_{1}
    (E1; \mathbf{k}, \hat{\boldsymbol{\varepsilon}})
    }
    {
    2
    \braket{I
    (
    E1';
    \mathbf{k}, 
    \hat{\boldsymbol{\varepsilon}}
    )
    }_{\mathrm{IC}}
    }\,.
    \label{eq:tnd_sum}
\end{align}
$\braket{I(E1'; \mathbf{k}, \hat{\boldsymbol{\varepsilon}})}_{\mathrm{IC}}$ and $\braket{t_{\partial k}(E1'; \mathbf{k}, \hat{\boldsymbol{\varepsilon}})}_{\mathrm{IC}}$ are independent of the light momentum $\boldsymbol{\kappa}$ because $\boldsymbol{\kappa}$-dependent terms (the third terms in Eqs.~(\ref{eq:MFPAD_E1'}) and~(\ref{eq:ted_gu})), which reflect the two-center interference phenomenon, cancel each other owing to the incoherent sum.
While the $\boldsymbol{\kappa}$-dependence stemmed from MFPAD, $I^{g/u}$, in $\braket{t_{\partial\kappa}(E1'; \mathbf{k}, \hat{\boldsymbol{\varepsilon}})}_{\mathrm{IC}}$ disappeared, $\braket{t_{\partial\kappa}(E1'; \mathbf{k}, \hat{\boldsymbol{\varepsilon}})}_{\mathrm{IC}}$ retaining the $\boldsymbol{\kappa}$-dependent factor unrelated the two-center interference phenomenon.
In the Direct Wave approximation discussed in subsection~\ref{sec:DW}, the discrepancy between $I_{1}$ and $I_{2}$ in the numerator of Eq.~(\ref{eq:tnd_sum}) becomes zero.
However, in this study, we did not neglect the scattering effects of photoelectrons; therefore, the numerator has a pure scattering contribution.
This is discussed in more detail in subsection~\ref{sec:MS}.

The following relationships between the differences in the MFPAD and photoemission time delay were obtained in the two opposite directions $\hat{\mathbf{k}}$ and $-\hat{\mathbf{k}}$:
\begin{align}
&   \braket{I
    (
    E1';
    \mathbf{k}, 
    \hat{\boldsymbol{\varepsilon}}
    )
    }_{\mathrm{IC}}
-
    \braket{I
    (
    E1';
    -\mathbf{k}, 
    \hat{\boldsymbol{\varepsilon}}
    )
    }_{\mathrm{IC}}
=
    0
    \label{eq:I-I=0}
\\
&   \braket{t
    (
    E1';
    \mathbf{k}, 
    \hat{\boldsymbol{\varepsilon}}, 
    \boldsymbol{\kappa}
    )
    }_{\mathrm{IC}}
-
    \braket{t
    (
    E1';
    -\mathbf{k}, 
    \hat{\boldsymbol{\varepsilon}}, 
    \boldsymbol{\kappa}
    )
    }_{\mathrm{IC}}
    \nonumber\\
&=
    2\,
    \braket{t_{\partial\kappa}
    (
    E1';
    \mathbf{k}, 
    \hat{\boldsymbol{\varepsilon}}, 
    \boldsymbol{\kappa}
    )
    }_{\mathrm{IC}}
    \,.
    \label{eq:t-t=tnd}
\end{align}
The proofs of these equations are provided in Appendix~\ref{proof_opposite}.
The relationship in Eq.~(\ref{eq:t-t=tnd}) allows for the extraction of the non-dipole effect.

\subsection{Decomposition of the photoemission time delay using Multiple Scattering theory with spherical wave correction}\label{sec:MS}

In our previous study on heteronuclear diatomic molecules, the photoemission time delay was decomposed into atomic, propagation and scattering components.
This decomposition was achieved by expanding the photoionization amplitude in the scattering order using Multiple Scattering theory.~\cite{Tamura2022}
We have adopted this approach to expand the photoionization amplitude $A_{i}$ in Eq.~(\ref{eq:amp_E1'}) leveraging the Muffin-tin approximation~\cite{Tamura2022} which utilizes spherically symmetric potentials.

The photoelectron wave excited at site $i$ is composed of a superposition of direct emission, $D_{i}$, single scattering $S_{i\rightarrow j}$, and multiple scattering waves $H_{i\rightleftharpoons j}$.
Consequently, the following expansion is obtained:
\begin{align}
&   
    A_{i}
    (E1; \mathbf{k}, \hat{\boldsymbol{\varepsilon}})
\xrightarrow[\substack{
    \mathrm{Multiple\,\,Scattering}
    }]{}
    -
    M^{\,1}_{L_{c}}(E1;k)\,\,
    \mathrm{i}\,\,
    \sqrt{\frac{k}{\pi}}\,\,
    T_{1}(k)
    \nonumber\\
&\times
    \left\{
    D_{i}
    (\mathbf{k}, 
    \hat{\boldsymbol{\varepsilon}})
    +
    S_{i\rightarrow j}
    (\mathbf{k}, 
    \hat{\boldsymbol{\varepsilon}})
    +
    H_{i\rightleftharpoons j}
    (\mathbf{k}, 
    \hat{\boldsymbol{\varepsilon}})
    \right\}
    \nonumber\\
&\hspace{5.2 cm}
    \equiv
    \overline{A}_{i}
    (E1; \mathbf{k}, \hat{\boldsymbol{\varepsilon}})
    \,,
\label{eq:A_MS}
\end{align}
where $M^{\,1}_{L_{c}}(E1;k)$ represents the transition matrix element describing the excitation from the core orbital of atom $i$ within $E1$ transition approximation and $T_{1}$ represents $T$-matrix (transition operator) for a partial wave of $l=1$ with the spherical potential of site $i$.
$D_{i}$ represents the contribution of photoelectrons directly emitted from site $i$.
$S_{i\rightarrow j}$ represents the single scattering contribution of photoelectrons emitted from site $i$ propagating to and scattered by site $j$.
$H_{i\rightleftharpoons j}$ represents the higher order scattering contribution of the photoelectrons between sites $i$ and $j$, emitted from either site $i$ or $j$.
$D_{i}$ and $S_{i\rightarrow j}$ can be expressed as follows:
\begin{align}
    D_{i}
    (\mathbf{k}, 
    \hat{\boldsymbol{\varepsilon}})
&=
    \cos\theta_{\hat{\boldsymbol{\varepsilon}},\hat{\mathbf{k}}\,
    }
    e^{-\mathrm{i}\mathbf{k}\cdot\mathbf{R}_{i\mathrm{O}}}
    \label{eq:Direct}
\end{align}
and
\begin{align}
    S_{i\rightarrow j}
    (\mathbf{k}, 
    \hat{\boldsymbol{\varepsilon}})
&=
    \cos\theta_{\hat{\boldsymbol{\varepsilon}},\hat{\mathbf{R}}_{ji}\,
    }
    \mathcal{G}(k,R_{ji})\,\,
    f^{\,\left(0\right)}(\mathbf{k},\mathbf{R}_{ji})\,\,
    e^{\mathrm{i}\mathbf{k}\cdot\mathbf{R}_{i\mathrm{O}}}
    \nonumber\\
&\hspace{-1.2 cm}+
    \sin\theta_{\hat{\boldsymbol{\varepsilon}},\hat{\mathbf{R}}_{ji}}\,
    \cos\phi_{
    \hat{\boldsymbol{\varepsilon}},
    \hat{\mathbf{R}}_{ji},
    \hat{\mathbf{k}}
    }\,\,
    \mathcal{G}(k,R_{ji})\,\,
    f^{\,\left(1\right)}(\mathbf{k},\mathbf{R}_{ji})\,\,
    e^{\mathrm{i}\mathbf{k}\cdot\mathbf{R}_{i\mathrm{O}}}
    \,,
    \label{eq:Single_Scattering}
\end{align}
where $\theta_{\,\hat{\mathbf{r}},\hat{\mathbf{r}}'}=\mathrm{arccos}(\hat{\mathbf{r}}\cdot\hat{\mathbf{r}}')$ and $\phi_{\hat{\boldsymbol{\varepsilon}},\hat{\mathbf{R}}_{ji},\hat{\mathbf{k}}}$ denotes the dihedral angle between the plane containing $\hat{\boldsymbol{\varepsilon}}$ and $\hat{\mathbf{R}}_{ji}$ and the plane containing $\hat{\mathbf{R}}_{ji}$ and $\hat{\mathbf{k}}$.
When $\hat{\boldsymbol{\varepsilon}}$, $\hat{\mathbf{R}}$ and $\hat{\mathbf{k}}$ are on the same plane, $\phi_{\hat{\boldsymbol{\varepsilon}},\hat{\mathbf{R}}_{ji},\hat{\mathbf{k}}}=0$.
This condition is employed for the numerical calculations described in section~\ref{results}.
$\mathbf{R}_{ji}$ represents the position vector from the center of site $i$ to $j$.
$\mathcal{G}(k,R)\equiv e^{\mathrm{i}kR}/R\,$ represents free electron green's function.
$f^{\left(0\right)}$ and $f^{\left(1\right)}$ represent the modified scattering amplitudes, including the spherical wave effects.
The details of the derivations are provided in Appendix~\ref{proof_sw} and Ref.~\cite{Natoli1989}.

We briefly discuss the contribution of the spherical wave components in Eq.~(\ref{eq:Single_Scattering}).
The first term in Eq.~(\ref{eq:Single_Scattering}) denotes the component of the excited photoelectron wave parallel to the molecular axis, as observed from the term $\cos\theta_{\hat{\boldsymbol{\varepsilon}},\hat{\mathbf{R}}_{ji}}=\hat{\boldsymbol{\varepsilon}}\cdot\hat{\mathbf{R}}_{ji}$.
In the Plane Wave limit $kR \gg 1$, the spherical correction factors $c_{l}(kR)\rightarrow1$ (Eq.~(\ref{eq:cl}) in Appendix~\ref{proof_sw}), comprises a single scattering contribution $S^{\mathrm{PW}}_{i\rightarrow j}$ in the Plane Wave approximation.
Unlike the first term, the second term in Eq.~(\ref{eq:Single_Scattering}) is a component of the photoelectron wave excited in the direction perpendicular to the molecular axis but propagates to the neighboring atom owing to spherical wave correction.
When the polarization vector $\hat{\boldsymbol{\varepsilon}}$ is perpendicular to the molecular axis, the propagation time of light between atoms reaches its maximum; however, the single scattering contribution $S_{i\rightarrow j}$ vanishes in the Plane Wave approximation because only the spherical contribution describes the propagation of the photoelectron wave along the molecular axis.
Spherical wave formalism was employed to consider this contribution.
\onecolumngrid 
Using the expansions described in Eqs.~(\ref{eq:A_MS})-(\ref{eq:Single_Scattering}), the $k$-partial derivative part of the photoemission time delay, $t_{\partial k}^{g/u}$, in Eq.~(\ref{eq:tnd_gu}) can be divided into three components,
\begin{align}
    t_{\partial k}^{g/u}
    (
    E1'; 
    \mathbf{k}, 
    \hat{\boldsymbol{\varepsilon}}, 
    \boldsymbol{\kappa}
    )
    \xrightarrow[\substack{
    \mathrm{Multiple\,\,Scattering}
    }]{}
t_{\mathrm{abs}}
    (E1;k)
+
    t^{g/u}_{\mathrm{path}}
    (
    E1'; 
    \mathbf{k}, 
    \hat{\boldsymbol{\varepsilon}}, 
    \boldsymbol{\kappa}
    )
+
    t^{g/u}_{\mathrm{sc}}
    (
    E1'; 
    \mathbf{k}, 
    \hat{\boldsymbol{\varepsilon}}, 
    \boldsymbol{\kappa}
    )
    \,,
    \label{eq:4t}
\end{align}
where
\begin{align}
    t_{\mathrm{abs}}(E1;k)
&\equiv
    \frac{1}{v_{\mathrm{g}}}
    \frac
    {\partial}
    {\partial k}
    \,
    \mathrm{arg}
    \left\{
    M^{\,1}_{L_{c}}(E1;k)\,\,
    T_{1}(k)
    \right\}\,,
    \label{eq:t_abs}
\end{align}
\begin{align}
    t^{g/u}_{\mathrm{path}}
    (
    E1'; 
    \mathbf{k}, 
    \hat{\boldsymbol{\varepsilon}}, 
    \boldsymbol{\kappa}
    )
&\equiv
    \left|\overline{A}^{g/u}(E1';\mathbf{k},\hat{\boldsymbol{\varepsilon}},\boldsymbol{\kappa})\right|^{-2}
\times
    \mathrm{Im}
    \Biggl\{
    \overline{A}^{g/u\,*}(E1';\mathbf{k},\hat{\boldsymbol{\varepsilon}},\boldsymbol{\kappa})\,\,
    \frac{-1}{\sqrt{2}}\,\,
    M^{\,1}_{L_{c}}(E1;k)\,\,
    \mathrm{i}\,\,
    \sqrt{\frac{k}{\pi}}\,\,
    T_{1}(k)
    \nonumber\\
&\hspace{-2.7 cm}\times
    \Biggl(
    \Biggl[
    e^{-\mathrm{i}\boldsymbol{\kappa}\cdot\frac{\mathbf{R}}{2}}\,\,
    \cos\theta_{\hat{\boldsymbol{\varepsilon}},\hat{\mathbf{k}}\,
    }
    \frac{1}{v_{\mathrm{g}}}
    \frac{\partial}{\partial k}
    \left(
    e^{-\mathrm{i}\mathbf{k}\cdot\frac{\mathbf{R}}{2}}
    \right)
\pm
    e^{\mathrm{i}\boldsymbol{\kappa}\cdot\frac{\mathbf{R}}{2}}\,\,
    \cos\theta_{\hat{\boldsymbol{\varepsilon}},\hat{\mathbf{k}}}\,
    \frac{1}{v_{\mathrm{g}}}
    \frac{\partial}{\partial k}
    \left(
    e^{\mathrm{i}\mathbf{k}\cdot\frac{\mathbf{R}}{2}}
    \right)
    \Biggr]
    \nonumber\\
&\hspace{-2.4 cm}
+
    \Biggl[
    e^{-\mathrm{i}\boldsymbol{\kappa}\cdot\frac{\mathbf{R}}{2}}
    \Biggl\{
    \cos\theta_{
    \hat{\boldsymbol{\varepsilon}},
    \hat{\mathbf{R}}
    }\,
    \frac{1}{v_{\mathrm{g}}}
    \frac{\partial}{\partial k}
    \left(
    \mathcal{G}(k,R)\,\,
    e^{-\mathrm{i}\mathbf{k}\cdot\frac{\mathbf{R}}{2}}
    \right)
    f^{\,\left(0\right)}(\mathbf{k},-\mathbf{R})
+
    \sin\theta_{\hat{\boldsymbol{\varepsilon}},\hat{\mathbf{R}}}\,
    \cos\phi_{
    \hat{\boldsymbol{\varepsilon}},
    \hat{\mathbf{R}},
    \hat{\mathbf{k}}
    }\,\,
    \frac{1}{v_{\mathrm{g}}}
    \frac{\partial}{\partial k}
    \left(
    \mathcal{G}(k,R)\,\,
    e^{-\mathrm{i}\mathbf{k}\cdot\frac{\mathbf{R}}{2}}
    \right)
    f^{\,\left(1\right)}(\mathbf{k},-\mathbf{R})
    \Biggr\}
    \nonumber\\
&\hspace{-1.9 cm}\pm
    e^{\mathrm{i}\boldsymbol{\kappa}\cdot\frac{\mathbf{R}}{2}}
    \Biggl\{
    \cos\theta_{
    \hat{\boldsymbol{\varepsilon}},
    -\hat{\mathbf{R}}
    }\,
    \frac{1}{v_{\mathrm{g}}}
    \frac{\partial}{\partial k}
    \left(
    \mathcal{G}(k,R)\,\,
    e^{\mathrm{i}\mathbf{k}\cdot\frac{\mathbf{R}}{2}}
    \right)
    f^{\left(0\right)}(\mathbf{k},\mathbf{R})
+
    \sin\theta_{\hat{\boldsymbol{\varepsilon}},-\hat{\mathbf{R}}}\,
    \cos\phi_{
    \hat{\boldsymbol{\varepsilon}},
    -\hat{\mathbf{R}},
    \hat{\mathbf{k}}
    }\,\,
    \frac{1}{v_{\mathrm{g}}}
    \frac{\partial}{\partial k}
    \left(
    \mathcal{G}(k,R)\,\,
    e^{\mathrm{i}\mathbf{k}\cdot\frac{\mathbf{R}}{2}}
    \right)
    f^{\left(1\right)}(\mathbf{k},\mathbf{R})
    \Biggr\}
    \Biggr]
    \nonumber\\
&\hspace{-2.2 cm}+
    \cdots
    \hspace{14.0 cm}
    \Biggr)
    \Biggr\}\,,
    \label{eq:t_path}
\end{align}
\onecolumngrid 
\begin{align}
    t^{g/u}_{\mathrm{sc}}
    (
    E1'; 
    \mathbf{k}, 
    \hat{\boldsymbol{\varepsilon}}, 
    \boldsymbol{\kappa}
    )
&\equiv
    \left|\overline{A}^{g/u}(E1';\mathbf{k},\hat{\boldsymbol{\varepsilon}},\boldsymbol{\kappa})\right|^{-2}
\times
    \mathrm{Im}
    \Biggl\{
    \overline{A}^{g/u\,*}(E1';\mathbf{k},\hat{\boldsymbol{\varepsilon}},\boldsymbol{\kappa})\,\,
    \frac{-1}{\sqrt{2}}\,\,
    M^{\,1}_{L_{c}}(E1;k)\,\,
    \mathrm{i}\,\,
    \sqrt{\frac{k}{\pi}}\,\,
    T_{1}(k)
    \nonumber\\
&\hspace{-2.7 cm}\times
    \Biggl(
    \Biggl[
    e^{-\mathrm{i}\boldsymbol{\kappa}\cdot\frac{\mathbf{R}}{2}}
    \Biggl\{
    \cos\theta_{
    \hat{\boldsymbol{\varepsilon}},
    \hat{\mathbf{R}}\,
    }
    \mathcal{G}(k,R)\,\,
    e^{-\mathrm{i}\mathbf{k}\cdot\frac{\mathbf{R}}{2}}
    \frac{1}{v_{\mathrm{g}}}
    \frac{\partial}{\partial k}
    \left(
    f^{\,\left(0\right)}(\mathbf{k},-\mathbf{R})
    \right)
+
    \sin\theta_{\hat{\boldsymbol{\varepsilon}},\hat{\mathbf{R}}}\,
    \cos\phi_{
    \hat{\boldsymbol{\varepsilon}},
    \hat{\mathbf{R}},
    \hat{\mathbf{k}}
    }\,\,
    \mathcal{G}(k,R)\,\,
    e^{-\mathrm{i}\mathbf{k}\cdot\frac{\mathbf{R}}{2}}
    \frac{1}{v_{\mathrm{g}}}
    \frac{\partial}{\partial k}
    \left(
    f^{\,\left(1\right)}(\mathbf{k},-\mathbf{R})
    \right)
    \Biggr\}
    \nonumber\\
&\hspace{-1.9 cm}\pm
    e^{\mathrm{i}\boldsymbol{\kappa}\cdot\frac{\mathbf{R}}{2}}
    \Biggl\{
    \cos\theta_{
    \hat{\boldsymbol{\varepsilon}},
    -\hat{\mathbf{R}}
    }\,
    \mathcal{G}(k,R)\,\,
    e^{\mathrm{i}\mathbf{k}\cdot\frac{\mathbf{R}}{2}}
    \frac{1}{v_{\mathrm{g}}}
    \frac{\partial}{\partial k}
    \left(
    f^{\left(0\right)}(\mathbf{k},\mathbf{R})
    \right)
+
    \sin\theta_{\hat{\boldsymbol{\varepsilon}},-\hat{\mathbf{R}}}\,
    \cos\phi_{
    \hat{\boldsymbol{\varepsilon}},
    -\hat{\mathbf{R}},
    \hat{\mathbf{k}}
    }\,\,
    \mathcal{G}(k,R)\,\,
    e^{\mathrm{i}\mathbf{k}\cdot\frac{\mathbf{R}}{2}}
    \frac{1}{v_{\mathrm{g}}}
    \frac{\partial}{\partial k}
    \left(
    f^{\left(1\right)}(\mathbf{k},\mathbf{R})
    \right)
    \Biggr\}
    \Biggr]
    \nonumber\\
&\hspace{-2.2 cm}+
    \cdots
    \hspace{14.0 cm}
    \Biggr)
    \Biggr\}
    \label{eq:t_sc}
\end{align}
and
\begin{align}
   \overline{A}^{g/u}
    (E1'; \mathbf{k}, \hat{\boldsymbol{\varepsilon}}, \boldsymbol{\kappa})
\equiv
    \frac{1}{\sqrt{2}}\,\,
    \Bigl(
    e^{-\mathrm{i}\boldsymbol{\kappa}\cdot\frac{\mathbf{R}}{2}}\,\,
    \overline{A}_{1}
    (E1; \mathbf{k}, \hat{\boldsymbol{\varepsilon}})
    \pm
    e^{\mathrm{i}\boldsymbol{\kappa}\cdot\frac{\mathbf{R}}{2}}\,\,
    \overline{A}_{2}
    (E1; \mathbf{k}, \hat{\boldsymbol{\varepsilon}})
    \Bigr)\,.
    \label{eq:amp_E1'_MT}
\end{align}
Here $t_{\mathrm{abs}}$ represents atomic time delay for absorbing atoms within $E1$ transition approximation.
This time delay is independent of the photoemission and light angles.
The time delays $t^{g/u}_{\mathrm{path}}$ and $t^{g/u}_{\mathrm{sc}}$ stem from the propagation of photoelectrons between atoms and the scattering of photoelectrons by atoms, respectively.
\subsection{Direct Wave contribution \texorpdfstring{$D_{i}$}{TEXT} for MFPAD and photoemission time delay}\label{sec:DW}

We adopted the simplest Direct Wave approximation, which ignored the scattering of photoelectrons and described the photoelectron wave function as a superposition of direct waves from atoms.
In this approximation, only the lowest term $D_{i}$ in Eq.~(\ref{eq:A_MS}) was considered.
Consequently, we obtained the following analytical expression for MFPAD:
\begin{align}
   I^{g/u}
    (E1'; \mathbf{k}, \hat{\boldsymbol{\varepsilon}}, \boldsymbol{\kappa})
\xrightarrow[\substack{
    \mathrm{Direct\,\,Wave}\\
    \mathrm{approximation}
    }]{}
    16\pi
    c
    \kappa
    k
    \left|\,
    M^{\,1}_{L_{c}}(k)\,\,
    T_{1}(k)\,\,
    \right|^{2}
    (
    \hat{\boldsymbol{\varepsilon}}
    \cdot
    \hat{\mathbf{k}}
    )^{2}
    \left[
    1
    \pm
    \cos
    \left\{
    \left(\mathbf{k}-\boldsymbol{\kappa}\right)
    \cdot
    \mathbf{R}
    \right\}
    \right]\,.
    \label{eq:MFPAD_DW}
\end{align}
This expression captures a two-center interference phenomenon shifted by $\boldsymbol{\kappa}\cdot\mathbf{R}$ owing to non-dipole effects, as reported by Ivanov \textit{et al.}~\cite{Ivanov2021}.

Conversely, the photoemission time delay within the Direct Wave approximation equals that of a single absorbing atom and does not demonstrate the properties of the molecular structure, such as angular dependence and the non-dipole effect:
\begin{align}
&
    t^{g/u}
    (
    E1'; 
    \mathbf{k}, 
    \hat{\boldsymbol{\varepsilon}}, 
    \boldsymbol{\kappa}
    )
\xrightarrow[\substack{
    \mathrm{Direct\,\,Wave}\\
    \mathrm{approximation}
    }]{}
    t_{\mathrm{abs}}(E1;k)\,.
    \label{eq:t_dw}
\end{align}
It is therefore necessary to describe the photoemission time delay in a model that considers the effects of scattering with the surrounding atoms unlike MFPADs.
\onecolumngrid 
\section{Results and Discussion}\label{results}


\subsection {Analytical expression in high energy region}

In our previous study of the photoemission time delay of heteronuclear diatomic molecules~\cite{Tamura2022}, we anticipate that $t_{\mathrm{path}}^{g/u\left(1\right)}$ and $\braket{t_{\mathrm{path}}^{\left(1\right)}}_{\mathrm{IC}}$ (the superscript $\left(1\right)$ denotes Single Scattering approximation) are dominant in the photoemission time delays $t^{g/u}$ and $\braket{t}_{\mathrm{IC}}$ in the high-energy region.
Finally, the following analytical expressions are obtained:
\begin{align}
    t_{\mathrm{path}}^{g/u\left(1\right)}
    (E1'; \mathbf{k}, \hat{\boldsymbol{\varepsilon}}, \boldsymbol{\kappa})
&=
    \frac
    {
    2
    \braket{I^{\left(1\right)}
    (E1; \mathbf{k}, \hat{\boldsymbol{\varepsilon}})
    }_{\mathrm{IC}}\,
    \braket{t^{\left(1\right)}_{\mathrm{path}}
    (E1; \mathbf{k}, \hat{\boldsymbol{\varepsilon}})
    }_{\mathrm{IC}}
    }
    {I^{g/u\left(1\right)}
    (E1'; \mathbf{k}, \hat{\boldsymbol{\varepsilon}}, \boldsymbol{\kappa})}
\pm
    \frac{
    \alpha(k)
    }
    {I^{g/u\left(1\right)}
    (E1'; \mathbf{k}, \hat{\boldsymbol{\varepsilon}}, \boldsymbol{\kappa})}
    \nonumber\\
&\hspace{0.5 cm}
    \times
    \frac{R}{v_{g}}
    \Bigl[
    (1+\hat{\mathbf{k}}\cdot\hat{\mathbf{R}})\,
    \mathrm{Re}\left\{
    D^{\,*}_{1}(\mathbf{k},\hat{\boldsymbol{\varepsilon}})\,\,
    S_{2\rightarrow 1}(\mathbf{k},\hat{\boldsymbol{\varepsilon}})
    \right\}
+
    (1-\hat{\mathbf{k}}\cdot\hat{\mathbf{R}})\,
    \mathrm{Re}\left\{
    D^{\,*}_{2}(\mathbf{k},\hat{\boldsymbol{\varepsilon}})\,\,
    S_{1\rightarrow 2}(\mathbf{k},\hat{\boldsymbol{\varepsilon}})
    \right\}
    \nonumber\\
&\hspace{8.5 cm}
+
    2\,
    \mathrm{Re}
    \left\{
    S^{\,*}_{1\rightarrow 2}(\mathbf{k},\hat{\boldsymbol{\varepsilon}})\,\,
    S_{2\rightarrow 1}(\mathbf{k},\hat{\boldsymbol{\varepsilon}})
    \right\}
    \Bigr]\,,
    \label{eq:t_path^{g/u(1)}}
\end{align}
and
\begin{align}
    \braket{t^{\left(1\right)}_{\mathrm{path}}
    (E1; \mathbf{k}, \hat{\boldsymbol{\varepsilon}})
    }_{\mathrm{IC}}
&=
    \frac
    {I^{\left(1\right)}_{1}(E1; \mathbf{k}, \hat{\boldsymbol{\varepsilon}})\,
    t_{\mathrm{path},1}^{\left(1\right)}
    (E1; \mathbf{k}, \hat{\boldsymbol{\varepsilon}})
+
    I^{\left(1\right)}_{2}(E1; \mathbf{k}, \hat{\boldsymbol{\varepsilon}})\,
    t_{\mathrm{path},2}^{\left(1\right)}
    (E1; \mathbf{k}, \hat{\boldsymbol{\varepsilon}})}
    {2\braket{I^{\left(1\right)}(E1; \mathbf{k}, \hat{\boldsymbol{\varepsilon}})}_{\mathrm{IC}}}
    \,,
    \label{eq:t_path^{sum(1)}}
\end{align}
where
\begin{align}  
    t^{(1)}_{\mathrm{path},i}
    (E1; \mathbf{k}, \hat{\boldsymbol{\varepsilon}})
&\equiv
    \frac{R}{v_{g}}
    \left(1-\hat{\mathbf{k}}\cdot\hat{\mathbf{R}}_{ji}\right)
    \frac{1}{2}
    \left(
    1
    -
    \frac{
    \left|
    D_{i}(\mathbf{k},\hat{\boldsymbol{\varepsilon}})
    \right|^{2}
    -
    \left|
    S_{i\rightarrow j}(\mathbf{k},\hat{\boldsymbol{\varepsilon}})
    \right|^{2}}
    {
    \left(\alpha(k)\right)^{-1}
    I^{\left(1\right)}_{i}
    (E1; \mathbf{k}, \hat{\boldsymbol{\varepsilon}})
    }
    \right)\,,
    \label{eq:t_path^{i(1)}}
\\
    I^{g/u\left(1\right)}
    (
    E1'; 
    \mathbf{k}, 
    \hat{\boldsymbol{\varepsilon}}, 
    \boldsymbol{\kappa}
    )
&=
    \braket{I^{\left(1\right)}
    (E1; \mathbf{k}, \hat{\boldsymbol{\varepsilon}})
    }_{\mathrm{IC}}
    \nonumber\\
&\hspace{0.5 cm}
\pm
    \alpha(k)
    \Biggl[
    \mathrm{Re}
    \Bigl\{
    e^{\mathrm{i}\boldsymbol{\kappa}\cdot\mathbf{R}}\,
    D^{\,*}_{1}
    (\mathbf{k}, \hat{\boldsymbol{\varepsilon}})\,
    D_{2}
    (\mathbf{k}, \hat{\boldsymbol{\varepsilon}})
    \Bigr\}
+
    \mathrm{Re}
    \Bigl\{
    e^{\mathrm{i}\boldsymbol{\kappa}\cdot\mathbf{R}}\,
    D^{\,*}_{1}
    (\mathbf{k}, \hat{\boldsymbol{\varepsilon}})\,
    S_{2\rightarrow 1}
    (\mathbf{k}, \hat{\boldsymbol{\varepsilon}})
    \Bigr\}
    \nonumber\\
&\hspace{1.5 cm}
+
    \mathrm{Re}
    \Bigl\{
    e^{\mathrm{i}\boldsymbol{\kappa}\cdot\mathbf{R}}\,
    S^{\,*}_{1\rightarrow 2}
    (\mathbf{k}, \hat{\boldsymbol{\varepsilon}})\,
    D_{2}
    (\mathbf{k}, \hat{\boldsymbol{\varepsilon}})
    \Bigr\}
+
    \mathrm{Re}
    \Bigl\{
    e^{\mathrm{i}\boldsymbol{\kappa}\cdot\mathbf{R}}\,
    S^{\,*}_{1\rightarrow 2}
    (\mathbf{k}, \hat{\boldsymbol{\varepsilon}})\,
    S_{2\rightarrow 1}
    (\mathbf{k}, \hat{\boldsymbol{\varepsilon}})
    \Bigr\}
    \Biggr]
    \,,
    \label{eq:I^{g/u(1)}}
\\
    \braket{I^{\left(1\right)}
    (E1; \mathbf{k}, \hat{\boldsymbol{\varepsilon}})
    }_{\mathrm{IC}}
&=
    \frac{1}{2}
    \left(
    I^{\left(1\right)}_{1}
    (E1; \mathbf{k}, \hat{\boldsymbol{\varepsilon}})
+
    I^{\left(1\right)}_{2}
    (E1; \mathbf{k}, \hat{\boldsymbol{\varepsilon}})
    \right)
    \,,
    \label{eq:I^{sum(1)}}
\\
    I^{\left(1\right)}_{i}
    (E1; \mathbf{k}, \hat{\boldsymbol{\varepsilon}})
&=
    \alpha(k)
    \Biggl[
    \left|
    D_{i}(\mathbf{k},\hat{\boldsymbol{\varepsilon}})
    \right|^{2}
    +
    \left|
    S_{i\rightarrow j}(\mathbf{k},\hat{\boldsymbol{\varepsilon}})
    \right|^{2}
+
    2
    \mathrm{Re}\left\{
    D^{\,*}_{i}(\mathbf{k},\hat{\boldsymbol{\varepsilon}})\,
    S_{i\rightarrow j}(\mathbf{k},\hat{\boldsymbol{\varepsilon}})
    \right\}
    \Biggl]\,,
    \label{eq:I^{i(1)}}
\\
    \alpha(k)
&\equiv
    8\pi ck\kappa\,\,
    \bigl|
    M^{\,1}_{L_{c}}(E1;k)\,\,
    T_{1}(k)
    \bigr|^{2}\,.
    \label{eq:C(k)}
\end{align}
\twocolumngrid 
\noindent
Here, $t^{(1)}_{\mathrm{path},i}$ is similar to that obtained for excitation from the localized core orbital at atomic site $i$ {of} heteronuclear diatomic molecules (Eq.~(20) from Ref.~\cite{Tamura2022}) except for spherical wave correction.

\begin{figure}[htbp!]
\includegraphics[width=0.35\textwidth]{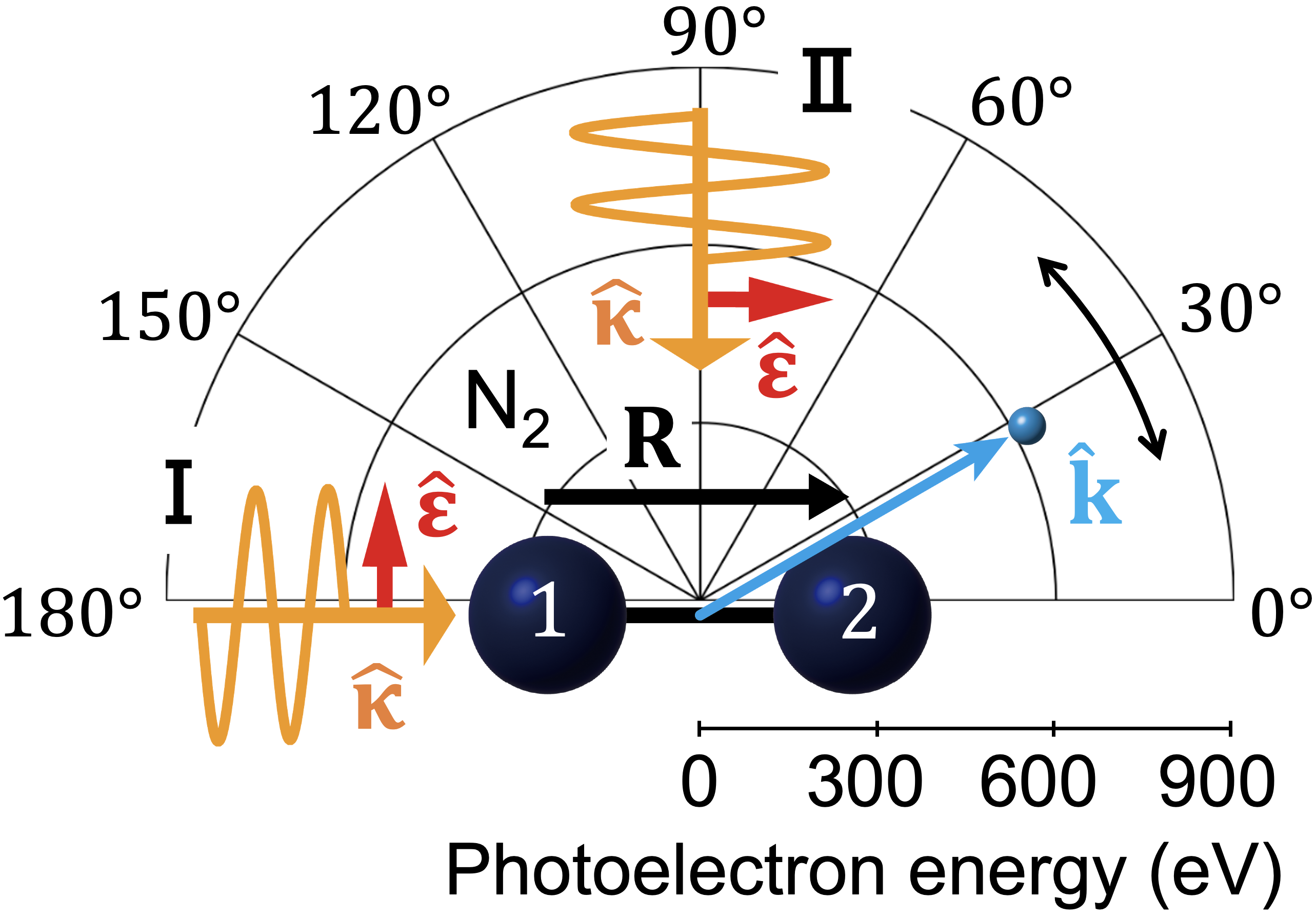}
\caption{Definition of the molecular frame for a nitrogen molecule with two distinct directions of incident light $\hat{\boldsymbol{\kappa}}$, which are (I) parallel and (II) perpendicular to the molecular axis $\hat{\mathbf{R}}$.
The photoemission direction $\hat{\mathbf{k}}$ and molecular axis $\hat{\mathbf{R}}$ are in the $\hat{\boldsymbol{\varepsilon}}$-$\hat{\boldsymbol{\kappa}}$ plane, where $\hat{\boldsymbol{\varepsilon}}$ denotes the polarization vector of incident light.
The energy of the photoelectron, $E_{\mathbf{k}}$, is determined by the radius of the semicircle, whereas the polar angles indicate the photoemission direction $\hat{\mathbf{k}}$ relative to the molecular axis $\hat{\mathbf{R}}$.
This polar plot is utilized in Figs.~\ref{fig:X}-\ref{fig:Y}.}
\label{fig:mol}
\end{figure}

\begin{figure*}[htbp!]
\includegraphics
[width=0.75\textwidth]{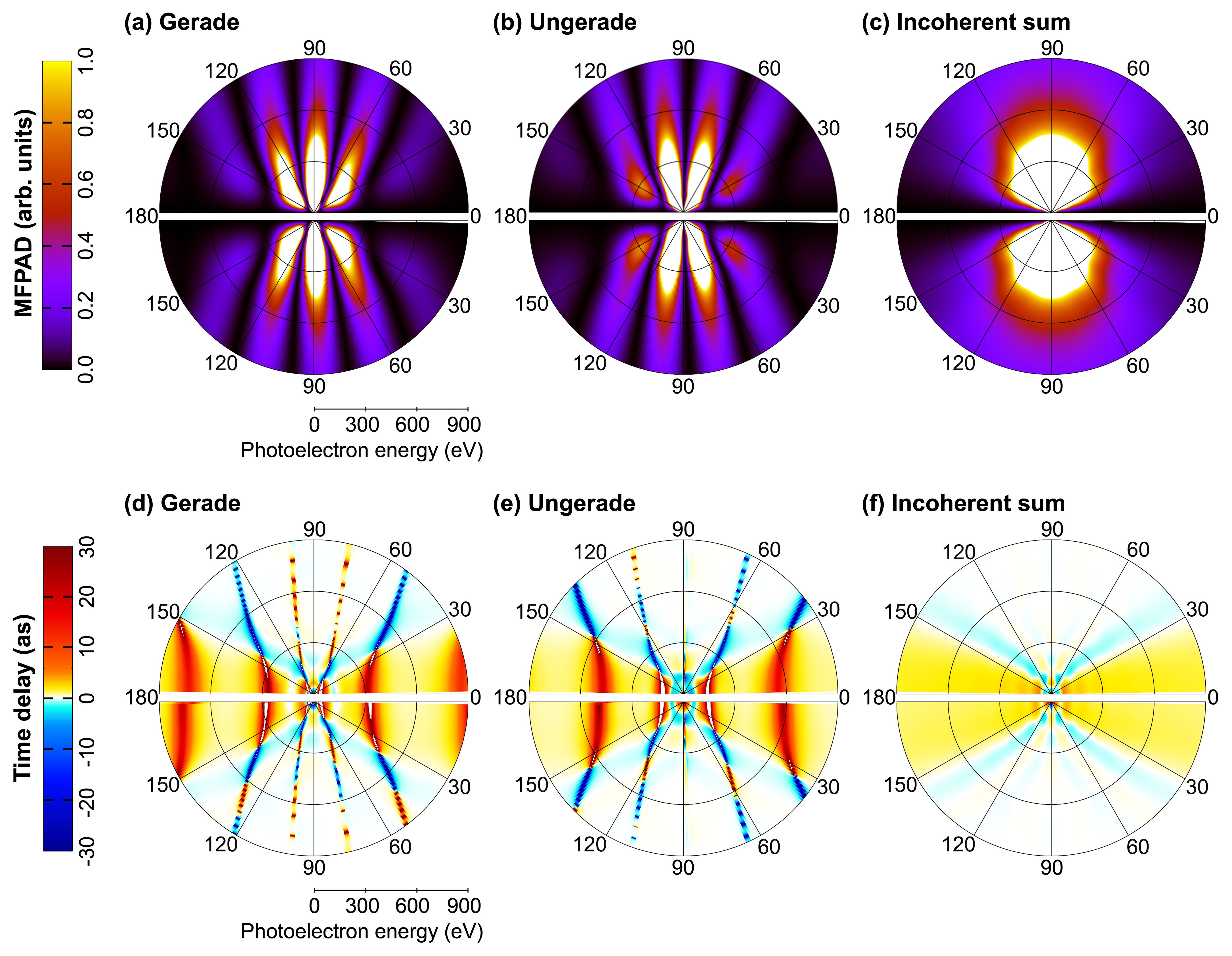}
\caption {Comparison of the numerical and analytical results for MFPAD and photoemission time delay from \textit{gerade} and \textit{ungerade} states of the 1$s$ core orbitals in nitrogen molecules in the coordinate system (I) in Fig.~\ref{fig:mol} ($\hat{\boldsymbol{\kappa}}\parallel\hat{\mathbf{R}}$) in terms of energy and angle dependence of the photoelectrons.
The upper semicircles in the polar plots represent the full multiple scattering numerical calculation results
of MFPAD for (a) \textit{gerade} and (b) \textit{ungerade} states, $I^{g/u}(E1'; \mathbf{k}, \hat{\boldsymbol{\varepsilon}}, \boldsymbol{\kappa})$, and (c) their incoherent sum $\braket{I(E1; \mathbf{k}, \hat{\boldsymbol{\varepsilon}})}_{\mathrm{IC}}$.
Difference in the photoemission time delay between nitrogen molecule and nitrogen atom for (d) \textit{gerade} and (e) \textit{ungerade} states, $t^{g/u}(E1'; \mathbf{k}, \hat{\boldsymbol{\varepsilon}}, \boldsymbol{\kappa})-t_{\mathrm{abs}}(E1;k)$, and (f) their incoherent sum, $\braket{t(E1; \mathbf{k}, \hat{\boldsymbol{\varepsilon}})}_{\mathrm{IC}}-t_{\mathrm{abs}}(E1;k)$.
The lower semicircles in the polar plots represent the analytical results of MFPAD for (a) \textit{gerade} and (b) \textit{ungerade} states, $I^{g/u\left(1\right)}(E1'; \mathbf{k}, \hat{\boldsymbol{\varepsilon}}, \boldsymbol{\kappa})$, in Eq.~(\ref{eq:I^{g/u(1)}}), and (c) their incoherent sum, $\braket{I^{\left(1\right)}(E1; \mathbf{k}, \hat{\boldsymbol{\varepsilon}})}_{\mathrm{IC}}$, in Eq.~(\ref{eq:I^{sum(1)}}).
Difference in the photoemission time delay between nitrogen molecules and atoms for (d) \textit{gerade} and (e) \textit{ungerade} states, $t_{\mathrm{path}}^{g/u\left(1\right)}(E1'; \mathbf{k}, \hat{\boldsymbol{\varepsilon}}, \boldsymbol{\kappa})$, in Eq.~(\ref{eq:t_path^{g/u(1)}}), and (f) their incoherent sum, $\braket{t^{\left(1\right)}_{\mathrm{path}}(E1; \mathbf{k}, \hat{\boldsymbol{\varepsilon}})}_{\mathrm{IC}}$, in Eq.~(\ref{eq:t_path^{sum(1)}}).
}
\label{fig:X}
\end{figure*}

Next, we present the numerically calculated results of the MFPAD and photoemission time delay of nitrogen molecules for two cases:
(I) the light incident direction $\hat{\boldsymbol{\kappa}}$ parallel to the molecular axis $\hat{\mathbf{R}}$ and 
(II) perpendicular to the molecular axis $\hat{\mathbf{R}}$ in the $\hat{\boldsymbol{\varepsilon}}$-$\hat{\boldsymbol{\kappa}}$ plane (refer to Fig.~\ref{fig:mol}).
The numerical outcomes for both cases were compared against the aforementioned analytical expressions to validate their accuracy (Figs.~\ref{fig:X} and~\ref{fig:Y}).
The following numerical results were obtained:
\begin{enumerate}[label=(\roman*), left=0pt]
    \item Multiple scattering calculations were performed for $A_{i}$ defined in Eq.~(\ref{eq:amp_i}) by MsSpec code~\cite{sebilleau2011msspec} within Muffin-tin approximation.
    \item The photoionization amplitude $A^{g/u}$ was calculated based on Eq.~(\ref{eq:amp_E1'}).
    \item The MFPAD and photoemission time delay was calculated in terms of Eqs.~(\ref{eq:MFPAD_after_E1'}) and~(\ref{eq:time_delay_after_E1'}), respectively.
    \item The incoherent sums were employed as the definitions in Eqs.~(\ref{eq:IS_MFPAD}) and~(\ref{eq:IS_time_delay}), respectively.
    \item $t_\mathrm{abs}$ was subtracted from each photoemission time delay.
\end{enumerate}

The quantities obtained from step (v) represent the difference in the photoemission time delay between the nitrogen molecule and nitrogen atom, that is $t^{g/u}-t_{\mathrm{abs}}$ and $\braket{t}_{\mathrm{IC}}-t_{\mathrm{abs}}$.
These values are measurable in the experiment~\cite{Heck2022}.
We assumed that the time delay caused by the Coulomb tail, which is the long-range Coulomb potential generated by the core hole, is effectively neutralized at step (v) for both the molecule and single atom utilized as the experimental reference.
Consequently, the effect of the Coulomb tail was disregarded in our numerical and analytical calculations.
The molecular potential was established by two touching atomic spheres and a null potential in the remaining space within the Muffin-tin approximation framework, as detailed in our prior research~\cite{Tamura2022}.
The molecular potential, which is a superposition of self-consistently calculated free-atom potentials for the ground state, was spherically averaged within each atomic sphere.
This does not significantly alter our discussion in high energy regions.

The core-level photoemission time delay can be obtained through angular streaking utilizing attosecond X-ray pulses from a free-electron laser.~\cite{Driver_Nature,Ji_arXiv}
In this experiment, an ionizing attosecond pulse was overlapped with a few-cycle infrared (IR) laser, and the measured photoemission time delay encompassed the time delay due to the streaking effects by the IR field.
By computing the time delay attributed to the IR laser, we can subtract the IR-related time delay and extract the one-photon photoemission time delay from the measured photoemission time delay~\cite{Driver_Nature,Ji_arXiv}.

A simple theoretical model enabled us to analyze the effect of charge migration in core-hole states, influenced by the probe IR field, on the photoemission time delay in the high energy photoelectron region.
To evaluate the time delay contribution, we examine two scenarios as examples: 1) point charges with $Z_{\mathrm{hole},1}=Z_{\mathrm{hole},2}=0.5$ were present at both atomic sites (core holes were delocalized throughout the diatomic molecule); 2) point charge with $Z_{\mathrm{hole},1}$ or $Z_{\mathrm{hole},2}=1$ was localized at one atomic site.
We utilize a Coulomb potential model to explore the variations in time delays.
The time delay $t^{c}_{l}$ for a partial wave of a photoelectron scattered by a Coulomb potential ($Z/k\ll 1$) from a point charge $Z$ with a sufficiently large momentum $k$ of the photoelectron can be expressed as follows (see Appendix~\ref{time_delay_coulomb}):
\begin{align}
    t^{c}_{l}(k,Z)
\sim
    \frac{1}{2}
    \frac{Z}{k^{3}}
    \left\{
    -
    \frac{1}{2}
    \frac{1}{l+1}
+
    \ln\left(l+1\right)
    \right\}\,.
\label{eq:time_coulombe}
\end{align}
Notably, $t^{c}_{l}$ is proportional to the point charge $Z$.
Considering the single scattering process, the photoelectron wave was scattered twice, that is, once at each atomic site, resulting in delayed twice by potentials with $Z=0.5$ (the time delay is proportional to $Z_{\mathrm{hole},1}+Z_{\mathrm{hole},2}=1$) in case 1) and delayed once by potential with $Z=1$ in case 2).
Therefore, in qualitative terms, the total time delays in the two distinct charge distribution scenarios yield approximately the same value.

\subsection{Light incident parallel to the molecular axis \texorpdfstring{$(\hat{\boldsymbol{\kappa}}\parallel\hat{\mathbf{R}})$}{TEXT}}

The influence of the non-dipole effect is most pronounced when the direction of light $\hat{\boldsymbol{\kappa}}$ is parallel to the molecular axis $\hat{\mathbf{R}}$ because the propagation time of light between atoms is at its maximum.
In this case, the analytical expression cannot be managed by the Plane Wave approximation.
In Eq.~(\ref{eq:Single_Scattering}) which represents the single scattering wave, the first term, primarily contributed by plane waves, diminishes, whereas the second term, mainly influenced by spherical waves, is maximized.

The analytical expressions for the MFPAD (Eqs.~(\ref{eq:I^{g/u(1)}}) and~(\ref{eq:I^{sum(1)}})) and photoemission time delay (Eqs.~(\ref{eq:t_path^{g/u(1)}}) and~(\ref{eq:t_path^{sum(1)}})) reproduce the numerical results in the high energy region well, by comparing between the numerical and analytical results, shown in the upper and lower semicircles of the polar plots in Fig.~\ref{fig:X}.
The MFPAD peaks exhibit tilts due to the non-dipole effect when distinguishing photoelectrons from \textit{gerade} and \textit{ungerade} states as shown in Fig.~\ref{fig:X}(a) and (b).
These tilt angles can be estimated using Eq.~(\ref{eq:MFPAD_DW}) based on the Direct Wave approximation.
As shown in Fig.~\ref{fig:X}(c), the incoherent sum of the MFPAD, which is independent of $\boldsymbol{\kappa}$ as in Eq.~(\ref{eq:I_sum}) does not exhibit this tilt.
The photoemission time delays for the \textit{gerade} and \textit{ungerade} states demonstrate significant peaks at energies and angles with minimal MFPADs (refer to Fig.~\ref{fig:X}(d) and (e)).
This phenomenon is attributed to the analytical expression for the $t_\text{path}$ of these states (Eq.~(\ref{eq:t_path^{g/u(1)}})), which is inversely proportional to the MFPAD.
The prominent peaks disappeared after the incoherent sum of the photoemission time delays was obtained, as shown in Fig.~\ref{fig:X}(f).

The symmetry of Fig.~\ref{fig:X}(f) with respect to the plane perpendicular to the direction of incident light $\hat{\boldsymbol{\kappa}}$ is noteworthy.
The lower semicircle of the analytical result ($\braket{t^{\left(1\right)}_{\mathrm{path}}(E1; \mathbf{k}, \hat{\boldsymbol{\varepsilon}})}_{\mathrm{IC}}$) is symmetric owing to the cancellation of $\hat{\boldsymbol{\kappa}}$-dependence in the incoherent sum.
However, the upper semicircle of the incoherent sum of the total photoemission time delays ($\braket{t(E1';\mathbf{k},\hat{\boldsymbol{\varepsilon}},\boldsymbol{\kappa})}_{\mathrm{IC}}$) is asymmetric.

\begin{figure}[htbp!]
\includegraphics[width=0.4\textwidth]{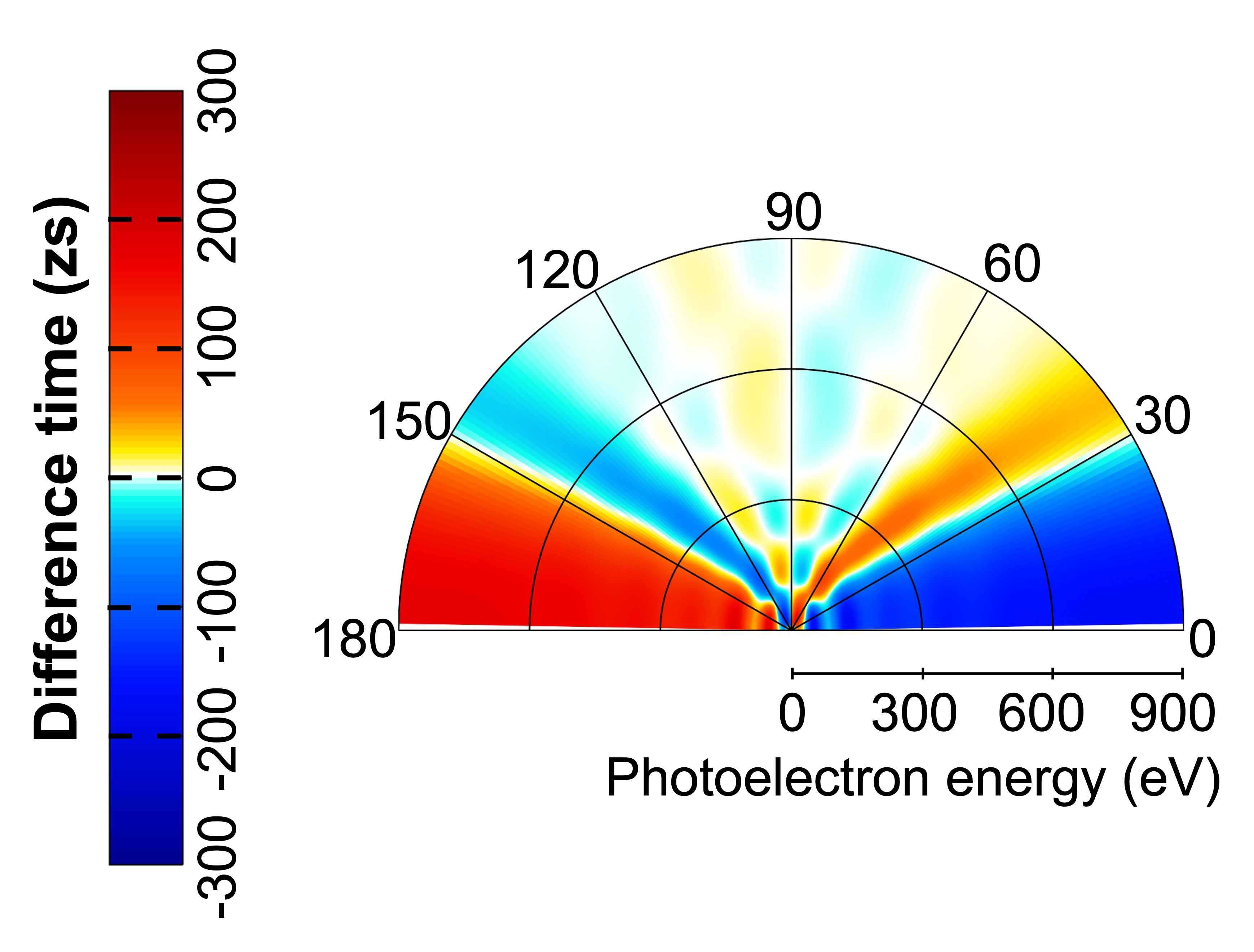}
\caption{Difference between the full multiple scattering numerical calculation results of the incoherent sum of photoemission time delays (upper semicircle in Fig.~\ref{fig:X}(f)) in any emission direction and the opposite direction, $\braket{t(E1';\mathbf{k},\hat{\boldsymbol{\varepsilon}},\boldsymbol{\kappa})}_{\mathrm{IC}}-\braket{t(E1';-\mathbf{k}, \hat{\boldsymbol{\varepsilon}}, \boldsymbol{\kappa})}_{\mathrm{IC}}$.}
\label{fig:asummetry}
\end{figure}

We analyzed the disparity in the upper semicircle as shown in Fig.~\ref{fig:X}(f) for each emission direction, denoted as $\hat{\mathbf{k}}$, and its opposite direction, $-\hat{\mathbf{k}}$.
Upon examination of the results shown in Fig.~\ref{fig:asummetry}, we observed minute variations that persisted at the zeptosecond level.
This zeptosecond asymmetry was estimated to be twice the $\kappa$-partial derivative component of the photoemission time delay, $2\braket{t_{\partial\kappa}(E1';\mathbf{k},\hat{\boldsymbol{\varepsilon}}, \boldsymbol{\kappa})}_{\mathrm{IC}}$, within our theoretical framework, as shown in Eq.~(\ref{eq:t-t=tnd}).

\subsection{Light incident perpendicular to the molecular axis \texorpdfstring{$(\hat{\boldsymbol{\kappa}}\perp\hat{\mathbf{R}})$}{TEXT}}

When the direction of light $\boldsymbol{\kappa}$ was perpendicular to the molecular axis $\hat{\mathbf{R}}$, the molecule-specific non-dipole effect disappeared.
This can be observed in Eq.~(\ref{eq:amp_E1'}):
the non-dipole factor $\exp{\pm\mathrm{i}\boldsymbol{\kappa}\cdot\mathbf{R}/2}$ becomes $1$.
Consequently, the forward and backward asymmetries in the MFPAD and photoemission time delay along the direction of incident light vanished.
In this case, the second term in Eq.~(\ref{eq:Single_Scattering}) (mostly spherical wave components) disappeared.

\begin{figure*}[htbp!]
\includegraphics
[width=0.75\textwidth]{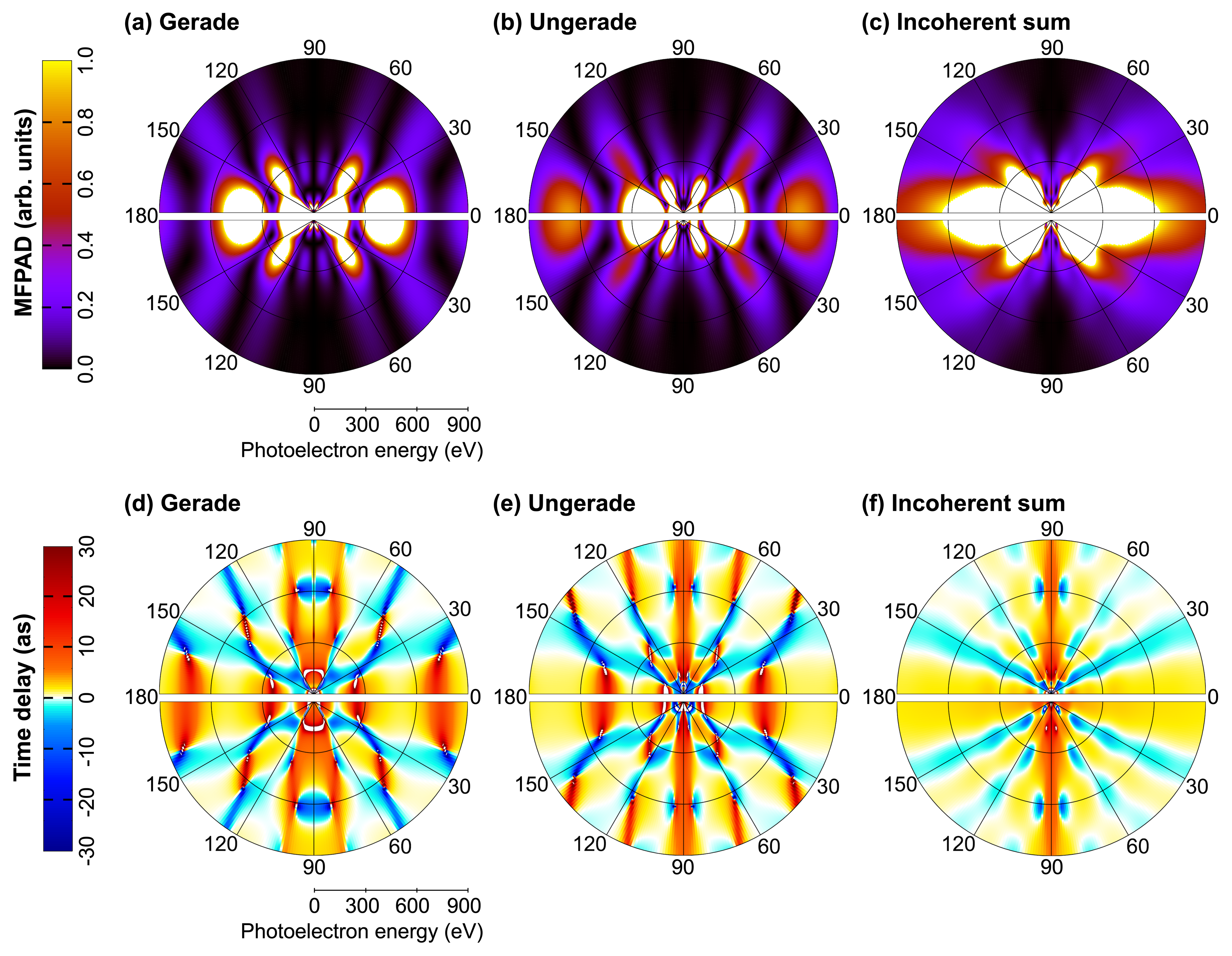}
\caption {
Comparison of numerical and analytical results for MFPAD and photoemission time delay from \textit{gerade} and \textit{ungerade} states of 1$s$ core orbitals in nitrogen molecules in the coordinate system (II) in Fig.~\ref{fig:mol} ($\hat{\boldsymbol{\kappa}}\perp\hat{\mathbf{R}}$) in terms of energy and angle dependence of the photoelectrons.
The upper semicircles in the polar plots represent the full multiple scattering numerical calculation results of MFPAD for (a) \textit{gerade} and (b) \textit{ungerade} states, $I^{g/u}(E1'; \mathbf{k}, \hat{\boldsymbol{\varepsilon}}, \boldsymbol{\kappa})$, and (c) their incoherent sum average, $\braket{I(E1; \mathbf{k}, \hat{\boldsymbol{\varepsilon}})}_{\mathrm{IC}}$.
Difference in the photoemission time delay between nitrogen molecules and atoms for (d) \textit{gerade} and (e) \textit{ungerade} states, $t^{g/u}(E1'; \mathbf{k}, \hat{\boldsymbol{\varepsilon}}, \boldsymbol{\kappa})-t_{\mathrm{abs}}(E1;k)$, and (f) their incoherent sum, $\braket{t(E1; \mathbf{k}, \hat{\boldsymbol{\varepsilon}})}_{\mathrm{IC}}-t_{\mathrm{abs}}(E1;k)$.
The lower semicircles in the polar plots denote the analytical results within Plane Wave approximation, Eq.~(\ref{eq:PWapproximation}), of MFPAD for (a) \textit{gerade} and (b) \textit{ungerade} states, $I^{g/u\left(1\right)}(E1'; \mathbf{k}, \hat{\boldsymbol{\varepsilon}}, \boldsymbol{\kappa})$, in Eq.~(\ref{eq:I^{g/u(1)}}), and (c) their incoherent sum, $\braket{I^{\left(1\right)}(E1; \mathbf{k}, \hat{\boldsymbol{\varepsilon}})}_{\mathrm{IC}}$, in Eq.~(\ref{eq:I^{sum(1)}}).
Difference in the photoemission time delay between nitrogen molecules and atoms for (d) \textit{gerade} and (e) \textit{ungerade} states, $t_{\mathrm{path}}^{g/u\left(1\right)}(E1'; \mathbf{k}, \hat{\boldsymbol{\varepsilon}}, \boldsymbol{\kappa})$, in Eq.~(\ref{eq:t_path^{g/u(1)}}), and (f) their incoherent sum, $\braket{t^{\left(1\right)}_{\mathrm{path}}(E1; \mathbf{k}, \hat{\boldsymbol{\varepsilon}})}_{\mathrm{IC}}$, in Eq.~(\ref{eq:t_path^{sum(1)}}).
}
\label{fig:Y}
\end{figure*}

In the high-energy region, the analytical expressions within the Plane Wave approximation (lower semicircle in Fig.~\ref{fig:Y}) can elucidate the behavior of the numerical results of the MFPAD and photoemission time delay (upper semicircle in Fig.~\ref{fig:Y}).

\section{Conclusion}

This study delved into the non-dipole effect observed in core-level photoemission of homonuclear diatomic molecules, utilizing a theoretical model beyond Electric Dipole approximation.
Initially, we elucidated the core-level MFPAD and photoemission time delay by employing the LCAO method.
Subsequently, we introduced the $E1'$ approximation to explore the non-dipole effect in the MFPAD and the photoemission time delay.
These approximations enabled the separation of a term $t_{\partial k}^{g/u}$, which was proportional to the attosecond-order interatomic propagation time of photoelectrons $R/v_{g}$, and a term $t_{\partial \kappa}^{g/u}$, proportional to the zeptosecond-order interatomic propagation time of light $R/c$.
The $t_{\partial \kappa}^{g/u}$ term provided zeptosecond temporal information specific to the photoemission time delay, a feature absent in the MFPAD.
Furthermore, we analytically derived fundamental relationships for $\braket{I}_{\mathrm{IC}}$ in Eq.~(\ref{eq:I-I=0}) and $\braket{t}_{\mathrm{IC}}$ in Eq.~(\ref{eq:t-t=tnd}).

Next, we divided $t_{\partial k}^{g/u}$ into an atomic time delay of a single absorbing atom $t_{\mathrm{abs}}$ and molecule-specific delays attributed to photoelectron propagation, $t_{\mathrm{path}}^{g/u}$, and scattering, $t_{\mathrm{sc}}^{g/u}$, by characterizing the photoelectron scattering state within the framework of Multiple Scattering theory.
We observed that molecule-specific delays $t_{\mathrm{path}}^{g/u}$, $t_{\mathrm{sc}}^{g/u}$ and $t_{\partial \kappa}^{g/u}$ vanished in the high energy region, where the Single Scattering approximation was valid.
Furthermore, we developed the analytical expressions 
(Eqs.~(\ref{eq:t_path^{g/u(1)}})–(\ref{eq:C(k)})) for the photoemission time delay in the high-energy region by considering up to single scattering.
Our analytical expressions were validated by comparing them with full multiple scattering numerical results for nitrogen molecules.

The photoemission time delay provided real-time insights into the zeptosecond-scale dynamics of light-molecule interactions, offering valuable information on molecular potential owing to the significant role of scattering contributions.
Our theoretical findings suggest that by measuring the photoemission time delay using advanced ultrashort light sources and detection technology, we can access information beyond MFPAD.
This study served as a foundation for theoretical investigations of photoionization phenomena in complex nano systems containing numerous identical atoms, such as large polyatomic molecules, liquids, and solids, enabling the exploration of photoemission time delays and non-dipole effects.

\section*{ACKNOWLEDGMENTS}
We extend our sincere gratitude to Mr. S. Sakamoto for the insightful discussions, and to Prof. Y. Hikosaka for conducting a thorough review.
We would like to extend our gratitude to Mr. K. Yoshikawa for generously providing valuable reference information.
This research was supported by JST and the establishment of university fellowships for the creation of science and technology innovation, Grant Number JPMJFS2115.
Additionally, K. Y. acknowledges the financial support received from JSPS KAKENHI Grant Number 19H05628.
This research was supported by JST, CREST Grant Number JPMJCR1861, Japan.

\appendix

\section{Treatment of initial orbitals and continuum states of homonuclear diatomic molecules and validity of \texorpdfstring{$E1'$}{TEXT} approximation}\label{derivation_E1'}
\begin{figure}[htbp!]
\includegraphics[width=0.3\textwidth]{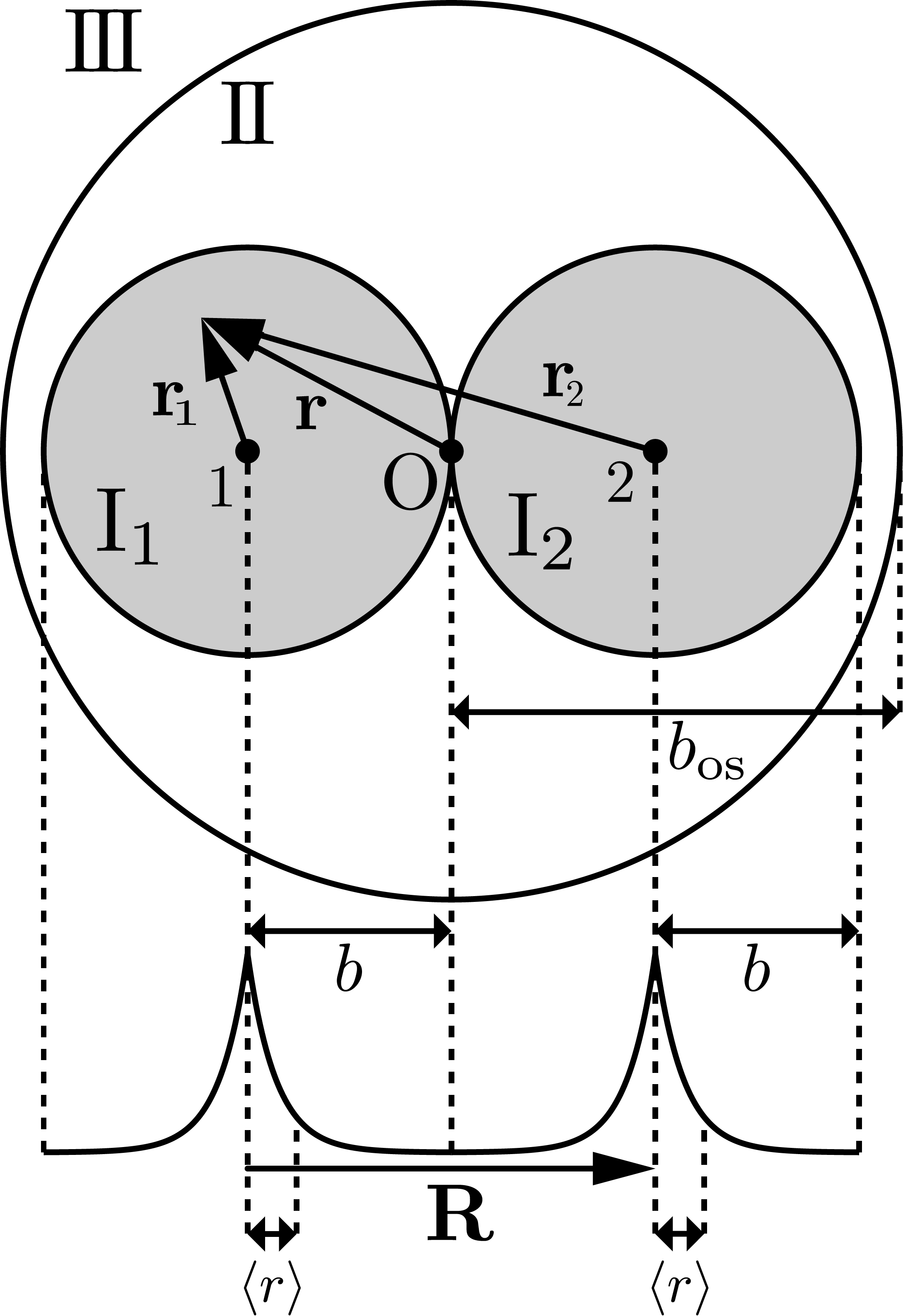}
\caption{Partitioning of the molecular space for homonuclear diatomic molecules by atomic spheres $1$ and $2$ (grey color).
These spheres are concentric with atoms (regions I$_1$ and I$_2$), and an outer sphere is centered at the origin O (the center of gravity of the molecule).
The interstitial space in the outer sphere was defined as region II and the external part of region II was defined as region III.
Here, $b$ represents a radius of regions I$_1$ and I$_2$, whereas $b_{\mathrm{os}}$ represents that of region II.
$\mathbf{r}$, $\mathbf{r}_{1}$ and $\mathbf{r}_{2}$ represent position vectors from O, sites $1$, and $2$, respectively.
$\mathbf{R}$ represents the vector from the site $1$ to $2$, and defines molecular axis.
Furthermore, we demonstrated a schematic of the wave function of the \textit{gerade} state within a linear combination of the 1$s$ hydrogenic atomic orbitals with atomic number $Z=7$.
This aimed to illustrate that the core orbital is tightly localized in the bond length $R$.
The mean radius of the localized core orbital $\braket{r}$ is sufficiently smaller than $R$.}
\label{fig:coordinate}
\end{figure}

We represent the \textit{gerade} and \textit{ungerade} states using a linear combination of atomic wave functions, which are spherically symmetric and confined in concentric atomic spheres with atoms $1$ and $2$, respectively.
These spheres have a radius of $b$ overlap with each other (refer to Fig.~\ref{fig:coordinate}):
\begin{align}
    \phi^{\,g/u}(\mathbf{r})
&=
    \frac{1}{\sqrt{2}}
    \Bigl[
    \Theta(r_{1}-b)\,\,
    \phi_{1,L_{c}}(r_{1})
    \pm
    \Theta(r_{2}-b)\,\,
    \phi_{2,L_{c}}(r_{2})
    \Bigr]\,,
    \label{eq:phi_gu}
\end{align}
where $\mathbf{r}$ represents a position vector from the origin $\mathrm{O}$ which is the center of gravity of the molecule.
$\Theta(x)$ represents a step function (if $x\leq0$ then $\Theta(x)=1$ else $\Theta(x)=0$).

The continuum wavefunction of photoelectron $\psi^{-}$ is partitioned into four regions in space, as follows~\cite{Dill1974}:
\begin{align}
    \psi^{-}(\mathbf{k}, \mathbf{r})
&=
    \sum_{j=1}^{2}\,\,
    \Theta(r_{j}-b)\,\,
    \psi^{-}_{j}(\mathbf{k}, \mathbf{r}_{j})
    \nonumber\\
&+
    \prod_{j=1}^{2}\,\,
    \bigl[1-\Theta(r_{j}-b)\bigr]\,\,
    \Theta(r-b_{\mathrm{os}})\,\,
    \psi^{-}_{\mathrm{I}\hspace{-1.2pt}\mathrm{I}}(\mathbf{k}, \mathbf{r})
    \nonumber\\
&+
    \bigl[1-\Theta(r-b_{\mathrm{os}})\bigr]\,\,
    \psi^{-}_{\mathrm{I}\hspace{-1.2pt}\mathrm{I}\hspace{-1.2pt}\mathrm{I}}(\mathbf{k}, \mathbf{r})\,,
    \label{eq:psi_-}
\end{align}
where $\psi^{-}_{j}$, $\psi^{-}_{\mathrm{I}\hspace{-1.2pt}\mathrm{I}}$ and $\psi^{-}_{\mathrm{I}\hspace{-1.2pt}\mathrm{I}\hspace{-1.2pt}\mathrm{I}}$ denote wavefunctions in regions $\mathrm{I}_{j}$, $\mathrm{I}\hspace{-1.2pt}\mathrm{I}$ and ${\mathrm{I}\hspace{-1.2pt}\mathrm{I}\hspace{-1.2pt}\mathrm{I}}$, respectively.
The region $\mathrm{I}_{j}$ is within the atomic sphere $j$, whereas the region $\mathrm{I}\hspace{-1.2pt} \mathrm{I}$ is the interstitial space inside the outer sphere with a radius $b_{\mathrm{os}}$ and is centered at the origin O.
The region ${\mathrm{I}\hspace{-1.2pt}\mathrm{I}\hspace{-1.2pt}\mathrm{I}}$ is located outside the outer sphere.
(refer to Fig.~\ref{fig:coordinate})

By substituting Eqs.~(\ref{eq:phi_gu}) and~(\ref{eq:psi_-}) into the left-hand side of Eq.~(\ref{eq:E1'_approximation}), the following equation was obtained:
\begin{align}
&
    \bra{\,\psi^{-}(\mathbf{k})\,}\,
    e^{\mathrm{i}\boldsymbol{\kappa}\cdot\mathbf{r}}\,\,
    \hat{\boldsymbol{\varepsilon}}
    \cdot 
    \mathbf{r}\,
    \ket{\,\phi^{\,g/u}}
    \nonumber\\
&=
    \frac{1}{\sqrt{2}}\,\,
    \int_{V_{1}}
    \mathrm{d}\mathbf{r}\,\,
    \psi^{-\,*}_{1}(\mathbf{k}, \mathbf{r}_{1})\,\,
    e^{\mathrm{i}\boldsymbol{\kappa}\cdot\mathbf{r}}\,\,
    \hat{\boldsymbol{\varepsilon}}
    \cdot 
    \mathbf{r}\,\,
    \phi_{1,L_{c}}(r_{1})
    \nonumber\\
&\pm
    \frac{1}{\sqrt{2}}\,\,
    \int_{V_{2}}
    \mathrm{d}\mathbf{r}\,\,
    \psi^{-\,*}_{2}(\mathbf{k}, \mathbf{r}_{2})\,\,
    e^{\mathrm{i}\boldsymbol{\kappa}\cdot\mathbf{r}}\,\,
    \hat{\boldsymbol{\varepsilon}}
    \cdot 
    \mathbf{r}\,\,
    \phi_{2,L_{c}}(r_{2})\,.
    \label{eq:amp_TCI}
\end{align}
The first and second terms on the right-hand side represent photoionization amplitudes that describe the excitation of electrons from the core orbitals of atoms $1$ and $2$, respectively, to the scattering states in the corresponding regions.

Next, we separate the light operators, $\exp\left(\mathrm{i}\boldsymbol{\kappa}\cdot\mathbf{r}\right)\hat{\boldsymbol{\varepsilon}}\cdot\mathbf{r}$, in Eq.~(\ref{eq:amp_TCI}), inducing multipoles into contributions inside and between the atoms.
Subsequently, we shifted the origin of the integrals to the center at each site $i$ and obtained the following equation:
\begin{align}
&
    \bra{\,\psi^{-}(\mathbf{k})\,}\,
    e^{\mathrm{i}\boldsymbol{\kappa}\cdot\mathbf{r}}\,\,
    \hat{\boldsymbol{\varepsilon}}
    \cdot 
    \mathbf{r}\,
    \ket{\,\phi^{\,g/u}}
    \nonumber\\
&=
    \frac{1}{\sqrt{2}}\,\,
    e^{\mathrm{i}\boldsymbol{\kappa}\cdot\mathbf{R}_{1\mathrm{O}}}
    \int_{V_{1}}
    \mathrm{d}\mathbf{r}_{1}\,\,
    \psi^{-\,*}_{1}(\mathbf{k}, \mathbf{r}_{1})\,\,
    e^{\mathrm{i}\boldsymbol{\kappa}\cdot\mathbf{r}_{1}}\,\,
    \hat{\boldsymbol{\varepsilon}}
    \cdot 
    \mathbf{r}_{1}\,\,
    \phi_{1,L_{c}}(r_{1})
    \nonumber\\
&+
    \frac{1}{\sqrt{2}}\,\,
    e^{\mathrm{i}\boldsymbol{\kappa}\cdot\mathbf{R}_{1\mathrm{O}}}\,\,
    \hat{\boldsymbol{\varepsilon}}
    \cdot 
    \mathbf{R}_{1\mathrm{O}}
    \int_{V_{1}}
    \mathrm{d}\mathbf{r}_{1}\,\,
    \psi^{-\,*}_{1}(\mathbf{k}, \mathbf{r}_{1})\,\,
    e^{\mathrm{i}\boldsymbol{\kappa}\cdot\mathbf{r}_{1}}\,\,
    \phi_{1,L_{c}}(r_{1})
    \nonumber\\
&\pm
    \frac{1}{\sqrt{2}}\,\,
    e^{\mathrm{i}\boldsymbol{\kappa}\cdot\mathbf{R}_{2\mathrm{O}}}
    \int_{V_{2}}
    \mathrm{d}\mathbf{r}_{2}\,\,
    \psi^{-\,*}_{2}(\mathbf{k}, \mathbf{r}_{2})\,\,
    e^{\mathrm{i}\boldsymbol{\kappa}\cdot\mathbf{r}_{2}}\,\,
    \hat{\boldsymbol{\varepsilon}}
    \cdot 
    \mathbf{r}_{2}\,\,
    \phi_{2,L_{c}}(r_{2})
    \nonumber\\
&\pm
    \frac{1}{\sqrt{2}}\,\,
    e^{\mathrm{i}\boldsymbol{\kappa}\cdot\mathbf{R}_{2\mathrm{O}}}\,\,
    \hat{\boldsymbol{\varepsilon}}
    \cdot 
    \mathbf{R}_{2\mathrm{O}}
    \int_{V_{2}}
    \mathrm{d}\mathbf{r}_{2}\,\,
    \psi^{-\,*}_{2}(\mathbf{k}, \mathbf{r}_{2})\,\,
    e^{\mathrm{i}\boldsymbol{\kappa}\cdot\mathbf{r}_{2}}\,\,
    \phi_{2,L_{c}}(r_{2})
    \label{eq:before_E1'}
    \\
&
    \xrightarrow[\substack{
    \mathrm{E1'\,\,approximation\,\,\left(e^{\mathrm{i}\boldsymbol{\kappa}\cdot\mathbf{r}_{1}},\,\,e^{\mathrm{i}\boldsymbol{\kappa}\cdot\mathbf{r}_{2}}\longrightarrow1\right)}
    }]{}
    \nonumber\\
&
    \frac{1}{\sqrt{2}}\,\,
    \Bigl(
    e^{-\mathrm{i}\boldsymbol{\kappa}\cdot\frac{\mathbf{R}}{2}}\,\,
    A_{1}
    (E1; \mathbf{k}, \hat{\boldsymbol{\varepsilon}})
    \pm
    e^{\mathrm{i}\boldsymbol{\kappa}\cdot\frac{\mathbf{R}}{2}}\,\,
    A_{2}
    (E1; \mathbf{k}, \hat{\boldsymbol{\varepsilon}})
    \Bigr)\,,
    \label{eq:Appendix_amp_E1'}
\end{align}
where the relationship $\mathbf{r}=\mathbf{r}_{i}+\mathbf{R}_{i\mathrm{O}}$ was utilized, with $\mathbf{R}_{i\mathrm{O}}$ representing a position vector from the origin $\mathrm{O}$ to the center position of atom $i$.
Therefore, $\mathbf{R}_{1\mathrm{O}}=-{\mathbf{R}}/{2}$ and $\mathbf{R}_{2\mathrm{O}}={\mathbf{R}}/{2}$.
(refer to Fig.~\ref{fig:coordinate}).
In the final step in Eq.~(\ref{eq:Appendix_amp_E1'}), we utilized $E1'$ approximation.
The $E1$ approximation was applied only for the integrands.
Upon applying the $E1'$ approximation, the orthogonality between the initial and final electronic wave functions was~\cite{taylor2006scattering}
\begin{align}
    \int_{V_{i}}
    \mathrm{d}\mathbf{r}_{i}\,\,
    \psi^{-\,*}_{i}(\mathbf{k}, \mathbf{r}_{i})\,\,
    \phi_{i,L_{c}}(\mathbf{r}_{i})
\sim
    \braket{\,
    \psi^{-}_{i}(\mathbf{k})\,|\,
    \phi_{i,L_{c}}
    }\,\,
\sim\,\,
    0\,.
\end{align}

Next, we estimated the validity of $E1'$ approximation using a hydrogen-like atom model.
In this model, the expected radius of the 1$s$ core orbital was $\braket{r}=3/\left(2Z\right)$ a.u., where $Z$ represents the atomic number.
The $\braket{r}$ for a nitrogen atom ($Z=7$) is expected to be $\braket{r}=0.21428$ a.u.
In this case, $E1$ approximation is valid in the integrals.
The non-dipole effect inside the atom~\cite{Amusia2020, Rezvan2022, Liang2024} may be ignored because the maximum absolute value for the quadrupole term is calculated as $\kappa\braket{r}\sim0.074796\ll1$ in multipole expansion of $\exp\left(\mathrm{i}\boldsymbol{\kappa}\cdot\mathbf{r}\right)\sim1+\mathrm{i}\boldsymbol{\kappa}\cdot\mathbf{r}+\cdots$ at a photoelectron energy of $900$ eV.
Conversely, the factor $\exp\left(\mathrm{i}\boldsymbol{\kappa}\cdot\mathbf{R}\right)$ is crucial in the non-dipole effect on photoemission time delay for nitrogen molecules owing to its bond length $R=2.0742$ a.u. and $\kappa R\sim 0.72400$.
Notably, the phase difference of the prefactors between $A_{1}$ and $A_{2}$ caused the non-dipole effect in the photoemission time delay.

\section{Proof of the relationships between two opposite directions of photoelectrons for the incoherent sums of the MFPAD and photoemission time delay}\label{proof_opposite}

\begin{figure}[htbp!]
\includegraphics[width=0.4\textwidth]{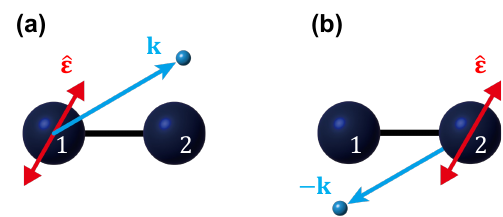}
\caption {
Photoelectron (a) from site $1$ to the direction $\hat{\mathbf{k}}$ and (b) from site $2$ to the direction $-\hat{\mathbf{k}}$ excited by the linearly polarized light with the polarization vector $\hat{\boldsymbol{\varepsilon}}$ for homonuclear diatomic molecules.
}
\label{fig:Fig6}
\end{figure}

The following basic relationships were obtained:
\begin{align}    
    I_{1}
    (E1;\mathbf{k}, \hat{\boldsymbol{\varepsilon}})
&=
    I_{2}
    (E1;-\mathbf{k}, \hat{\boldsymbol{\varepsilon}})\,,
    \label{eq:I1(k)=I2(-k)}
\\
    t_{1}
    (E1;\mathbf{k}, \hat{\boldsymbol{\varepsilon}})
&=
    t_{2}
    (E1;-\mathbf{k}, \hat{\boldsymbol{\varepsilon}})\,.
    \label{eq:t1(k)=t2(-k)}
\end{align}
These equations can be validated from a geometrical perspective.
By utilizing the symmetry of MFPAD, $I_{1}(E1;\mathbf{k}, \hat{\boldsymbol{\varepsilon}})$, and $t_{1}(E1;\mathbf{k}, \hat{\boldsymbol{\varepsilon}})$ (Fig.~(\ref{fig:Fig6}) (a)) can be rotated $180$ degrees in the $\hat{\boldsymbol{\varepsilon}}$-$\hat{\mathbf{k}}$ plane and function as a mirror reflection in relation to the same plane.
This rotation and reflection result in the functions coinciding with $I_{2}(E1;-\mathbf{k}, \hat{\boldsymbol{\varepsilon}})$ and $t_{2}(E1;-\mathbf{k}, \hat{\boldsymbol{\varepsilon}})$ (Fig.~(\ref{fig:Fig6}) (b)), respectively.
Eqs.~(\ref{eq:I-I=0}) and~(\ref{eq:t-t=tnd}) are proven by Eq.~(\ref{eq:teq_sum}) with Eqs.~(\ref{eq:I1(k)=I2(-k)}) and~(\ref{eq:t1(k)=t2(-k)}).

\section{Spherical wave correction formalism of Multiple Scattering theory}
\label{proof_sw}
Here, we summarize the components utilized to derive Eq.~(\ref{eq:A_MS}), using the spherical wave correction formalism in Refs.~\cite{Rehr1990, Natoli1989}.
The matrix representation of the free electron Green's function was expanded using the method of Rehr and Albers~\cite{Rehr1990} as follows:
\begin{align}
    G_{LL'}^{\,ij}
&\equiv
    -4\pi
    \mathrm{i}k\,\,
    \sum_{L''}\,\,
    \mathrm{i}^{\,l+l''-l'}\,\,
    C(L''L'L)\,\,
    h_{l''}(kR_{ij})\,\,
    \mathcal{Y}_{L''}(\hat{\mathbf{R}}_{ij})
    \\
&=
    \sum_{\mu}\,\,
    \sum_{\nu=0}^{\left|\mu\right|}\,\,
    \Theta(l-\left|\mu\right|)\,\,
    \mathrm{i}^{l}\,\,
    R_{\,m\mu}^{\,l}(\Omega_{\hat{\mathbf{R}}_{ij}}^{-1})\,\,
    \tilde{\gamma}_{\,\left|\mu\right|\nu}^{\,l}(kR_{ij})\,\,
    \nonumber\\
&\hspace{0.0 cm}\times
    \mathcal{G}_{ij}(k)\,\,
    \gamma_{\,\left|\mu\right|\nu}^{\,l'}(kR_{ij})\,\,
    R_{\,\mu m'}^{\,l'}(\Omega_{\hat{\mathbf{R}}_{ij}})\,\,
    \mathrm{i}^{-l'}\,\,
    \Theta(l'-\left|\mu\right|)\,,
\end{align}
where
\begin{align}
    \tilde{\gamma}_{\,\mu\nu}^{\,l}(kR)
&=
    \frac{2l'+1}{N_{l\mu}}\,\,
    \frac
    {\mathrm{d}^{\left(\nu\right)}C_{\,l}(z)}
    {\mathrm{d}z^{\left(\nu\right)}}\,\,
    \frac
    {z^{\,\nu}}
    {\nu!}\,,
\\
    \gamma_{\,\mu\nu}^{\,l'}(kR)
&=
    -
    N_{l'\mu}\,\,
    \left(-1\right)^{\mu}\,\,
    \frac{\mathrm{d}^{\left(\mu+\nu\right)}C_{\,l'}(z)}{\mathrm{d}z^{\left(\mu+\nu\right)}}\,\,
    \frac
    {z^{\mu+\nu}}
    {\left(\mu+\nu\right)!}\,,
\\
    N_{\,l\mu}
&\equiv
    \left(-1\right)^{\mu}
    \sqrt{\frac{2l+1}{4\pi}}\,\,
    \sqrt{\frac
    {\left(l-\mu\right)!}
    {\left(l+\mu\right)!}\,,
    }
\\
    c_{l}(kR)
&\equiv
    \sum_{n=0}^{l}
    \frac{\mathrm{i}^{n}}{n!\left(2kR\right)^{n}}
    \frac{\left(l+n\right)!}{\left(l-n\right)!}\,.
    \label{eq:cl}
\end{align}
$\mathcal{Y}_{L}(\hat{\mathbf{r}})$ represents real spherical harmonics for the angular momentum quantum number $L=(l, m)$, $l$ represents the angular azimuthal quantum number, $m$ represents the magnetic quantum number, and $C(L''L'L)$ represents Gaunt integral for real spherical harmonics.
$h_{l}$ represents Hankel function of the first kind.
$R_{m\mu}^{l}$ represents the rotation matrix element, $\Omega_{\hat{\mathbf{r}}}$ denotes the rotation that takes the vector $\hat{\mathbf{r}}$ along the $z$-axis, and $\Omega_{\hat{\mathbf{r}}}^{-1}$ represents the reverse rotation of it.
$\mu\geq0$, $z=1/\left(\mathrm{i}kR\right)$ and $C_{l}(z)=c_{l}(kR)$.
$c_{l}(kR)$ represents the factor of the spherical Hankel function, expressed as follows:
$h^{+}_{l}(kR)=\mathrm{i}^{-l}\exp(\mathrm{i}kR)/\left(kR\right)c_{l}(kR)$.

Within the framework of the Multiple Scattering theory in the Muffin-tin approximation~\cite{Tamura2022}, the core-level photoionization amplitude in Eq.~(\ref{eq:amp_i}) can be expanded as follows:
\begin{align}
&
    A_{i}
    (E1; \mathbf{k}, \hat{\boldsymbol{\varepsilon}})
    \xrightarrow[\substack{\mathrm{Muffin-tin}, \\
    \mathrm{Series\,\,Expansion}}]{}\,
    \nonumber\\
&
    \sqrt{\frac{k}{\pi}}\,\,
    \frac{4\pi}{3}\,\,
    \sqrt{\frac{1}{4\pi}}\,\,
    M_{1}^{L_{c}}(E1;k)
    \nonumber\\
&\times
    \Biggl\{
    \sum_{n}\,\,
    \mathcal{Y}_{1n}(\hat{\boldsymbol{\varepsilon}})\,\,
    T_{1}^{i}(k)\,\,
    \mathrm{i}^{1}\,\,
    \mathcal{Y}_{1n}(-\hat{\mathbf{k}})\,\,
    e^{-\mathrm{i}\mathbf{k}\cdot\mathbf{R}_{iO}}
    \nonumber\\
&\hspace{0.5 cm}+
    \sum_{j\left(\neq i\right)}\,\,
    \sum_{L_{1}}\,\,
    \sum_{n}\,\,
    \mathcal{Y}_{1n}(\hat{\boldsymbol{\varepsilon}})\,\,
    T^{\,i}_{1}(k)\,\,
    \nonumber\\
&\hspace{1.0 cm}\times
    G^{\,ij}_{1nL_{1}}\,\,
    T^{\,j}_{l_{1}}(k)\,\,
    \mathrm{i}^{l_{1}}\,\,
    \mathcal{Y}_{L_{1}}(-\hat{\mathbf{k}})\,\,
    e^{-\mathrm{i}\mathbf{k}\cdot\mathbf{R}_{jO}}
    \nonumber\\
&\hspace{0.5 cm}+
    \sum_{k\left(\neq i\right)}\,\,
    \sum_{j\left(\neq k\right)}\,\,
    \sum_{L_{1}L_{2}}\,\,
    \sum_{n}\,\,
    \mathcal{Y}_{1n}(\hat{\boldsymbol{\varepsilon}})\,\,
    T^{\,i}_{1}(k)\,\,
    \nonumber\\
&\hspace{1.0 cm}\times
    G^{\,ik}_{1nL_{1}}\,\,
    T^{\,k}_{l_{1}}(k)\,\,
    \nonumber\\
&\hspace{1.0 cm}\times
    G^{\,kj}_{L_{1}L_{2}}\,\,
    T^{\,j}_{l_{2}}(k)\,\,
    \mathrm{i}^{l_{2}}\,\,
    \mathcal{Y}_{L_{2}}(-\hat{\mathbf{k}})\,\,
    e^{-\mathrm{i}\mathbf{k}\cdot\mathbf{R}_{jO}}
    +\cdots\Biggr\}\,,
    \label{eq:appendix_amp_MS_1}
\end{align}
where 
\begin{align}
    M_{L_{c}}^{l}(E1;k)
&\equiv
    \int_{V_{i}}\,\,
    \mathrm{d}r_{i}\,\,
    R^{\,i\,*}_{\,l}(k, r_{i})\,\,
    r_{i}^{3}\,\,
    R^{\,i}_{L_{c}}(r_{i})
\end{align}
denotes the transition matrix and $T_{l}^{\,i}$ denotes the $T$-matrix element at site $i$ with $l$.
The major contributions of the direct and single scattering waves in Eq.~(\ref{eq:appendix_amp_MS_1}) are derived as follows:
\begin{align}
&
    D_{i}
    (\mathbf{k}, 
    \hat{\boldsymbol{\varepsilon}})
    =
    \cos\theta_{\hat{\boldsymbol{\varepsilon}},\hat{\mathbf{k}}}\,
    e^{-\mathrm{i}\mathbf{k}\cdot\mathbf{R}_{iO}}
\end{align}
and
\begin{align}
&
    S_{i\rightarrow j}
    (\mathbf{k}, 
    \hat{\boldsymbol{\varepsilon}})
    =
    \cos\theta_{\hat{\boldsymbol{\varepsilon}},\hat{\mathbf{R}}_{ji}}\,
    \mathcal{G}(k,R_{ji})\,\,
    f^{\,\left(0\right)}(\mathbf{k},\mathbf{R}_{ji})\,\,
    e^{\mathrm{i}\mathbf{k}\cdot\mathbf{R}_{iO}}
    \nonumber\\
&+
    \sin\theta_{\hat{\boldsymbol{\varepsilon}},\hat{\mathbf{R}}_{ji}}\,
    \cos\phi_{
    \hat{\boldsymbol{\varepsilon}},
    \hat{\mathbf{R}}_{ji},
    \hat{\mathbf{k}}
    }\,\,
    \mathcal{G}(k,R_{ji})\,\,
    f^{\,\left(1\right)}(\mathbf{k},\mathbf{R}_{ji})\,\,
    e^{\mathrm{i}\mathbf{k}\cdot\mathbf{R}_{iO}}\,,
\end{align}
where
\begin{align}
    \cos\phi_{
    \hat{\boldsymbol{\varepsilon}},
    \hat{\mathbf{R}}_{ji},
    \hat{\mathbf{k}}
    }
&=
    \frac{
    \cos\theta_{
    \hat{\boldsymbol{\varepsilon}},
    \hat{\mathbf{k}}
    }
    -
    \cos\theta_{
    \hat{\boldsymbol{\varepsilon}},\hat{\mathbf{R}}_{ji}
    }
    \cos\theta_{
    \hat{\mathbf{R}}_{ji},\hat{\mathbf{k}}
    }
    }
    {
    \sin\theta_{\hat{\boldsymbol{\varepsilon}},\hat{\mathbf{R}}_{ji}}\,
    \sin\theta_{\hat{\mathbf{R}}_{ji},\hat{\mathbf{k}}}
    }\,,
\\
    f^{\left(0\right)}(\mathbf{k},\mathbf{R})
&=
    \sum_{l=0}^{\infty}\,\,
    \left(2l+1\right)
    T_{\,l}(k)\,\,
    P_{l}(\hat{\mathbf{k}}\cdot\hat{\mathbf{R}})
    \nonumber\\
&\hspace{1.0 cm}\times
    \left(
    c_{\,l}(kR)
    -
    \frac{c_{\,l}(kR)}{\mathrm{i}kR}
    +
    \frac{1}{\mathrm{i}}
    \frac{\mathrm{d}c_{\,l}(kR)}{\mathrm{d}\left(kR\right)}
    \right)\,,
\\
    f^{\left(1\right)}(\mathbf{k},\mathbf{R})
&=
    \sum_{l=1}^{\infty}\,\,
    \left(2l+1\right)
    T_{\,l}(k)\,\,
    P_{l}^{\,1}
    (\hat{\mathbf{k}}\cdot\hat{\mathbf{R}})\,\,
    \frac{c_{\,l}(kR)}{\mathrm{i}kR}\,.
\end{align}
Here $\theta_{\,\hat{\mathbf{r}},\hat{\mathbf{r}}'}=\mathrm{arccos}(\hat{\mathbf{r}}\cdot\hat{\mathbf{r}}')$ and $\phi_{\hat{\boldsymbol{\varepsilon}},\hat{\mathbf{R}}_{ji},\hat{\mathbf{k}}}$ represents the dihedral angle between the plane containing $\hat{\boldsymbol{\varepsilon}}$ and $\hat{\mathbf{R}}_{ji}$ and that containing $\hat{\mathbf{R}}_{ji}$ and $\hat{\mathbf{k}}$.
$P_{l}$ and $P_{l}^{\,1}$ represent the Legendre function and Associated Legendre polynomials, respectively.~\cite{Natoli1989}

The analytical expressions within Plane Wave approximation were obtained by neglecting the second term of the single scattering wave $S_{i\rightarrow j}$ in Eq.~(\ref{eq:Single_Scattering}) and replacing the modified scattering amplitude as follows:
\begin{align}
&
    S_{i\rightarrow j}(\mathbf{k},\mathbf{R}_{ji})\,\,
\xrightarrow[\substack{
    \mathrm{Plane\,\,Wave}\\
    \mathrm{approximation}
    }]{}
    \nonumber\\
&
    \hat{\boldsymbol{\varepsilon}}
    \cdot
    \hat{\mathbf{R}}_{ji}\,\,
    \mathcal{G}(k,R_{ji})\,\,
    f(k,\theta_{\hat{\mathbf{R}}_{ji},\hat{\mathbf{k}}})\,\,
    e^{\mathrm{i}\mathbf{k}\cdot\mathbf{R}_{i\mathrm{O}}}
    \nonumber\\
&   
    \hspace{4.0 cm}
    \equiv
    S^{\mathrm{PW}}_{i\rightarrow j}
    (\mathbf{k},\mathbf{R}_{ji})\,,
    \label{eq:PWapproximation}
\end{align}
where
\begin{align}
    f(k,\theta_{\hat{\mathbf{R}}_{ji},\hat{\mathbf{k}}})
=
    \sum_{l=0}^{\infty}\,\,
    \left(2l+1\right)
    T_{\,l}(k)\,\,
    P_{l}(\hat{\mathbf{k}}\cdot\hat{\mathbf{R}}_{ji})
    \,,
    \label{eq:PWapproximation_for_f}
\end{align}
represents the ordinary scattering amplitude with a spherical symmetry potential.
\section{A time delay of partial waves due to scattering by a Coulomb potential in the high energy region}\label{time_delay_coulomb}
The phase shift for a partial wave with angular momentum quantum number $l$ owing to scattering by the Coulomb potential $Z/r$ is expressed according to Ref.~\cite{Barata2011} as follows:
\begin{align}
    \sigma_{l}(k,Z)
&\sim
    \left(l+\frac{1}{2}\right)
    \arctan\left(\frac{\eta}{l+1}\right)
-
    \eta
    \nonumber\\
&\hspace{0.0cm}+
    \frac{1}{2}
    \eta
    \ln\left\{\left(l+1\right)^{2}+\eta^{2}\right\}
    \nonumber\\
&\hspace{0.0cm}-
    \frac{1}{12}
    \frac{\eta}{\left(l+1\right)^{2}+\eta^{2}}\,,
\end{align}
where $\eta=-Z/k$ for an electron momentum $k$.
The time delay $t^{c}_{l}$ corresponding to this phase shift is obtained using the $k$-derivative with the group velocity $v_{g}$:
\begin{align}
    t^{c}_{l}(k,Z)
&\equiv
    \frac{1}{v_{g}}
    \frac{\mathrm{d}}{\mathrm{d}k}
    \sigma_{l}(k,Z)
\\
&=
    \frac{1}{2}
    \frac{Z}{k^{3}}
    \frac{\mathrm{d}}{\mathrm{d}\eta}
    \sigma_{l}(k,Z)
\\
&\sim
    \frac{1}{2}
    \frac{Z}{k^{3}}
    \Biggl[
    \frac{1}{\left(l+1\right)^{2}}
    \left\{
    \left(l+\frac{1}{2}\right)
    \left(l+1\right)
+
    \eta^{2}
-
    \frac{1}{12}
    \right\}
\nonumber\\
&+
    \frac{\eta^{2}}{6}
    \frac{1}{\left\{\left(l+1\right)^{2}+\eta^{2}\right\}^{2}}
+
    \frac{1}{2}
    \ln\left\{\left(l+1\right)^{2}+\eta^{2}\right\}
    \Biggr]
\\
&\sim
    \frac{1}{2}
    \frac{Z}{k^{3}}
    \left\{
    -
    \frac{1}{2}
    \frac{1}{l+1}
+
    \ln\left(l+1\right)
    \right\}\,.
\end{align}
In the final equation, we overlooked the second and higher-order terms of $\eta$ because $\eta=Z/k\ll 1$ was considered.

\providecommand{\noopsort}[1]{}\providecommand{\singleletter}[1]{#1}%

\end{document}